\documentclass[journal]{IEEEtran}

\usepackage{cite}
\usepackage{array}
\usepackage{color}
\usepackage{float}
\usepackage[cmex10]{amsmath}
\usepackage{nomencl}						
\usepackage[normalem]{ulem}			
\usepackage{diagbox}						
\usepackage{slashbox}						
\usepackage{booktabs} 
\usepackage{colortbl}						
\usepackage{multirow}						
\usepackage{tabularx}
\usepackage{placeins}
\usepackage{algorithm}
\usepackage[noend]{algpseudocode}
\usepackage{siunitx}
\usepackage{gensymb}

\ifCLASSINFOpdf
  \usepackage[pdftex]{graphicx}
	\graphicspath{{./figure/}}
  \DeclareGraphicsExtensions{.pdf,.jpeg,.png}
\else
  \usepackage[dvips]{graphicx}
  \graphicspath{{./figure/}}
  \DeclareGraphicsExtensions{.eps}
\fi

\ifCLASSOPTIONcompsoc
  \usepackage[caption=false,font=normalsize,labelfont=sf,textfont=sf]{subfig}
\else
  \usepackage[caption=false,font=footnotesize]{subfig}
\fi

\newcommand{\figref}[1]{\textcolor{black}{Fig.~\ref{#1}}}

\begin{document}
\bstctlcite{IEEEexample:BSTcontrol}
\setcounter{page}{1}

\onecolumn
\vspace*{0pt}
This work will be submitted to the IEEE for possible publication. Copyright may be transferred without notice, after which this version may no longer be accessible.
\newpage
\twocolumn

\title{Ideally-Smooth Transition between Grid-Forming and Grid-Following Inverters based on State Mapping Method}
\author{Zhenshuai Liu, Yitong Li \IEEEmembership{Member, IEEE}, Zirui Wang, Jiashuo Gu, ,Yao Qin, Jinjun Liu, \IEEEmembership{Fellow, IEEE}}

\ifCLASSOPTIONpeerreview
	\maketitle 
\else
	\maketitle
\fi


\begin{abstract}
There has been widespread global increasing use of renewable energy sources, which are usually connected to the electricity grids via power electronic inverters. Traditionally, these inverter-based resources operate in either grid-forming (GFM) or grid-following (GFL) mode. But more recently, the need of switching between these two modes are glowingly required because of the complex operation scenarios of systems such as source-side limitations, grid-side services, fault disturbances, etc. However, due to the differences between GFM and GFL modes, a direct switching between them would lead to large oscillations or even instability of inverters. Therefore, in this paper, a method called \textit{state mapping method} for analyzing the switching transient and designing the switching control is proposed. Based on this method, an ideally-smooth transition between GFM and GFL can be achieved. The effectiveness of the proposed method is verified by both the theoretical analysis and experiment tests.
\end{abstract}




\section{Introduction} \label{Section:Introduction}

As the penetration of renewable energy sources continues to rise, power electronic inverters have been playing a crucial role in the safe and stable operation of modern power systems. Conventional grid-following control (GFL) typically relies on a phase-locked loop (PLL) for synchronization to achieve fast and accurate current regulation. Grid-forming control (GFM) operates with voltage-source behavior, establishing voltage and frequency references to actively support the system \cite{li2022revisiting} \cite{rocabert2012control}. Given that power systems with high renewable penetration must adapt to diverse operational requirements, the transition between GFL and GFL mode is becoming more and more important. Consequently, achieving a smooth transition between GFL and GFM has become an essential requirement \cite{li2021impedance}.

To reduce voltage and current transients caused by abrupt mode transitions, many studies have focused on the control architectures and control strategies. A universal controller for voltage-source converters is proposed, in which GFL and GFM behaviors are achieved by adjusting the outer-loop reference generation \cite{harnefors2020universal}. While a further unified control scheme based on indirect current control is proposed \cite{kwon2021unified} \cite{liu2013unified}. These methods employ a unified mathematical framework to achieve seamless transitions between grid-connected and island mode, thereby eliminating the transients. This often compromises dynamic performance, making it challenging to simultaneously achieve the fast response of GFL and the robust synchronization of GFM. In \cite{li2021impedance}, an impedance-adaptive dual-mode control strategy is proposed to handle large SCR variations by control modes or adjusting parameters. A control strategy between GFL and GFM is proposed based on a multi-inverter system using the D-partition method \cite{Li2021The}. In \cite{ding2024novel}, a switchable GFL/GFM control structure is designed with optimized logic. For special operating conditions such as unbalanced grid voltages, a seamless transition method is proposed to ensure stable operation \cite{kim2025seamless}. The Kalman-filter-based observer is used to extract phase information and apply feedforward compensation during transition in \cite{cheng2024smooth}. In \cite{gao2024seamless}, a steady-state matching method is proposed to match the system parameters before and after the transition. A method for the between GFL and GFM of an inverter is proposed by correcting the voltage amplitude and phase reference \cite{Wang2018Enhanced}.

At the system level, such as in renewable energy farms, some scholars have also conducted researches on the smooth transition. In \cite{arafat2014effective}, it introduced a dispatch unit in PV farms to buffer the energy by temporarily providing additional power and voltage regulation that smooths the transition. A graph-theory-based stability analysis method for the transition is proposed by analyzing the coupling relationships among multiple nodes in a DC field station system \cite{Jiang2026Graph}. A smooth transition control strategy based on a PV station is proposed by analyzing the key disturbance factors to achieve a higher-density photovoltaic system \cite{Wang2023Smooth}. A fault ride-through control method with a negative-sequence current suppression strategy is proposed to achieve smooth transition between GFL and GFM under grid fault conditions \cite{Wang2025Flexible}. 

There have been many studies on the stability analysis of the system before and after the transition. In \cite{Ai2024Extension}, a new extension of GFM is proposed by analyzing the droop characteristics of GFM, which forms the voltage but does not support the frequency. To model and analyze the inverter more intuitively, a method named the impedance circuit model, which does not rely on experience, is proposed \cite{Li2021ImpedanceModel}. Using Lyapunov’s Direct Method, the nonlinear transient stability analysis of PLL-Based inverter is conducted \cite{Mansour2022Nonlinear}. The stability of GFL and GFM is analyzed through large-signal modeling in \cite{Fu2021Large}. In \cite{Dong2015Analysis}, a large-signal model incorporating PLL dynamics is developed, and it is shown that the effect of the grid can be represented as positive feedback. In \cite{Huang2017Bifurcation}, a bifurcation is identified for GFL during a fault event. This bifurcation can lead to loss of synchronization, from which recovery may be impossible.

Although existing research has made significant progress in the smooth transition between GFL and GFM through controller design or topology innovation, the underlying mechanism for achieving an ideally-smooth transition under arbitrary scenarios is still missing. Many studies have noted that a smooth transition can be achieved by changing parameters without altering the system control architecture, but few studies provide a rationale for doing this. The fundamental principle for achieving a smooth transition between GFL and GFM has not yet been fully revealed. To address this gap, this paper models the transition between GFL and GFM based on the concepts of the equilibrium point and the domain of attraction, investigates the mechanism for achieving smooth transition, and proposes a systematic method for analyzing the mode transition and designing the controllers. Here are the specific contributions:

(a) By analyzing the differences between GFL and GFM, and based on the necessary conditions for achieving the smooth transition between GFL and GFM, a general state space representation is proposed that provides a foundation for analyzing the transition between GFL and GFM.

(b) The Lyapunov's theorems are applied to explore the equilibrium points and the domain of attraction before and after the transition. Based on these, the transition process between GFL and GFM is modeled and analyzed, and an analytical method named the \textit{state mapping method} is proposed to achieve an ideally-smooth transition between GFL and GFM.

(c) Based on theoretical analysis, controllers are designed, namely the synchronization controller and the amplitude controller, which realize the ideally-smooth transition between GFL and GFM in practice.

The specific organization of the paper is as follows:  In Section \ref{Analysis of Mode Transition Transients}, by proposing the unified state space representations, a method called the state mapping method for analyzing the transition between GFL and GFM is proposed. In Section \ref{Control Design for Smooth Mode Transition}, based on Section \ref{Modeling Mode Transition}, a new controller and the simulation results of mode transition will be discussed carefully in this section. In Section \ref{Section:Experimental Results}, experimental results on a single-inverter-infinite-bus system are presented. Section \ref{Section:Conclusion} contains the conclusions.

\begin{figure}[th!]
\centering
\includegraphics[scale=0.65]{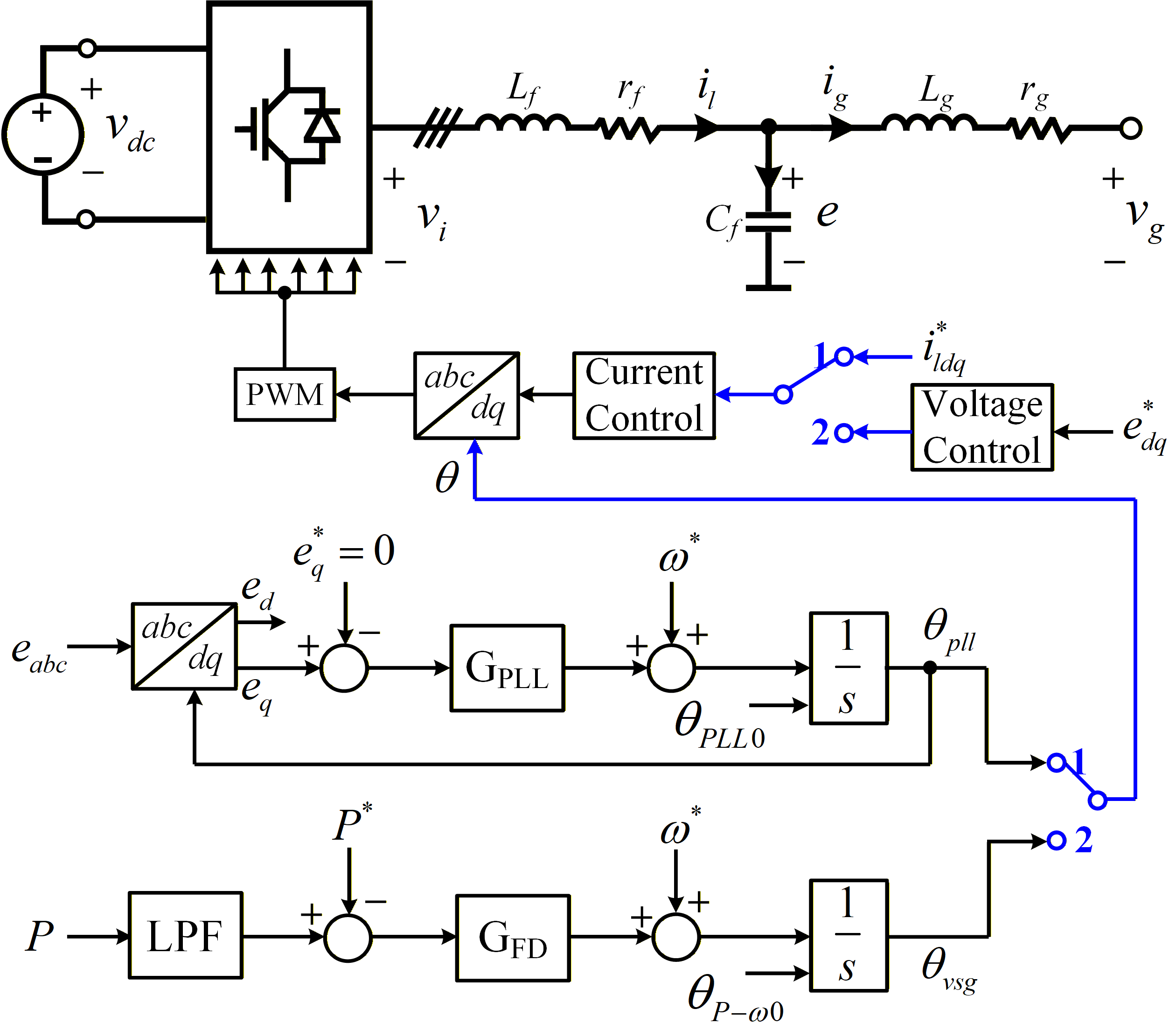}
\caption{Modeling of GFL and GFM used for studying smooth transition.}
\label{Fig:totalworkl}
\end{figure}


\section{Analysis of Mode Transition Transients} \label{Analysis of Mode Transition Transients}

\subsection{State Space Modeling}\label{Modeling Mode Transition}

The control structure of the inverter used for smooth mode transition is shown in \figref{Fig:totalworkl}. The grid-following control(GFL) shown in the figure includes an SRF-PLL and a current loop, while the grid-forming control(GFM) includes a $P-\omega$ droop controller, a current loop, and a voltage loop. The state-space representations of GFL and GFM are shown in (\ref{GFL}) and (\ref{GFM}), respectively.

\begin{equation}\label{GFL}
\begin{cases}
v_c = \frac{1}{C_f}\int(i_l-i_g)dt \\[6pt]
i_g = \frac{1}{L_g}\int(v_c-v_g-R_gi_g)dt \\[6pt]
i_l = \frac{1}{L_f}\int(v_i-v_c-R_fi_l)dt \\[6pt]
\theta_{PLL} = \int \omega_{PLL}dt + \theta_0 \\[6pt]
\omega_{PLL} = \omega_0 + k_pv_q+k_i\int v_qdt \\[6pt]
v^*_{idq} = K_{pi}(i^*_{ldq}-i_{ldq})+K_{ii}\int(i^*_{ldq}-i_{ldq})dt+v_{i0}
\end{cases}
\end{equation}

\begin{equation}\label{GFM}
\begin{cases}
v_c = \frac{1}{C_f}\int(i_l-i_g)dt \\[6pt]
i_g = \frac{1}{L_g}\int(v_c-v_g-R_gi_g)dt \\[6pt]
i_l = \frac{1}{L_f}\int(v_i-v_c-R_fi_l)dt \\[6pt]
\theta_{P-\omega} = \int (\omega_0+m_p(P^*-p\times LPF))dt + \theta_0 \\[6pt]
i^*_{ldq} = K_{pv}(v^*_{dq}-v_{dq})+K_{iv}\int(v^*_{dq}-v_{dq})dt+i_{i0} \\[6pt]
v^*_{idq} = K_{pi}(i^*_{ldq}-i_{ldq})+K_{ii}\int(i^*_{ldq}-i_{ldq})dt+v_{i0}
\end{cases}
\end{equation}

By comparing (\ref{GFL}) and (\ref{GFM}), it is easy to find that there are many similarities in state-space representations between GFL and GFM, such as the presence of $LCL$ filter parameters, the same input parameters and so on. It is greatly helpful to construct a unified state-space representations in identifying the factors that enable smooth transitions between GFL and GFM, as shown in (\ref{AB}) and (\ref{xc1}).

\begin{equation}\label{AB}
\begin{bmatrix}
\dot{x}_{c1} \\
\dot{x}_{c2} \\
\dot{x}_{phy}
\end{bmatrix}
=
\begin{bmatrix}
A_{11} & A_{12} & A_{13} \\
A_{21} & A_{22} & A_{23} \\
A_{31} & A_{32} & A_{33} 
\end{bmatrix}
\begin{bmatrix}
x_{c1} \\
x_{c2} \\
x_{phy}
\end{bmatrix}
+
\begin{bmatrix}
B_{11} & B_{12} \\
B_{21} & B_{22} \\
B_{31} & B_{32}
\end{bmatrix}
\begin{bmatrix}
u_{1} \\
u_2
\end{bmatrix}
\end{equation}
where
\begin{equation}\label{xc1}
\begin{cases}
x_{c1} =
\begin{bmatrix}
x_{c1,GFL} & x_{c1,GFM}
\end{bmatrix}
^T \\[6pt]
x_{c2} = i_{idq} \\[6pt]
x_{phy} = 
\begin{bmatrix}
v_c & i_g & i_l
\end{bmatrix}
^T \\[6pt]
u_1 = 
\begin{bmatrix}
u_{1,GFL} & u_{1,GFM}
\end{bmatrix}
^T \\[6pt]
u_2 = 
\begin{bmatrix}
v_g & v^*_{idq}
\end{bmatrix}
^T \\[6pt]
\end{cases}
\end{equation}

\begin{figure*}[th!]
\centering
\includegraphics[scale=0.6]{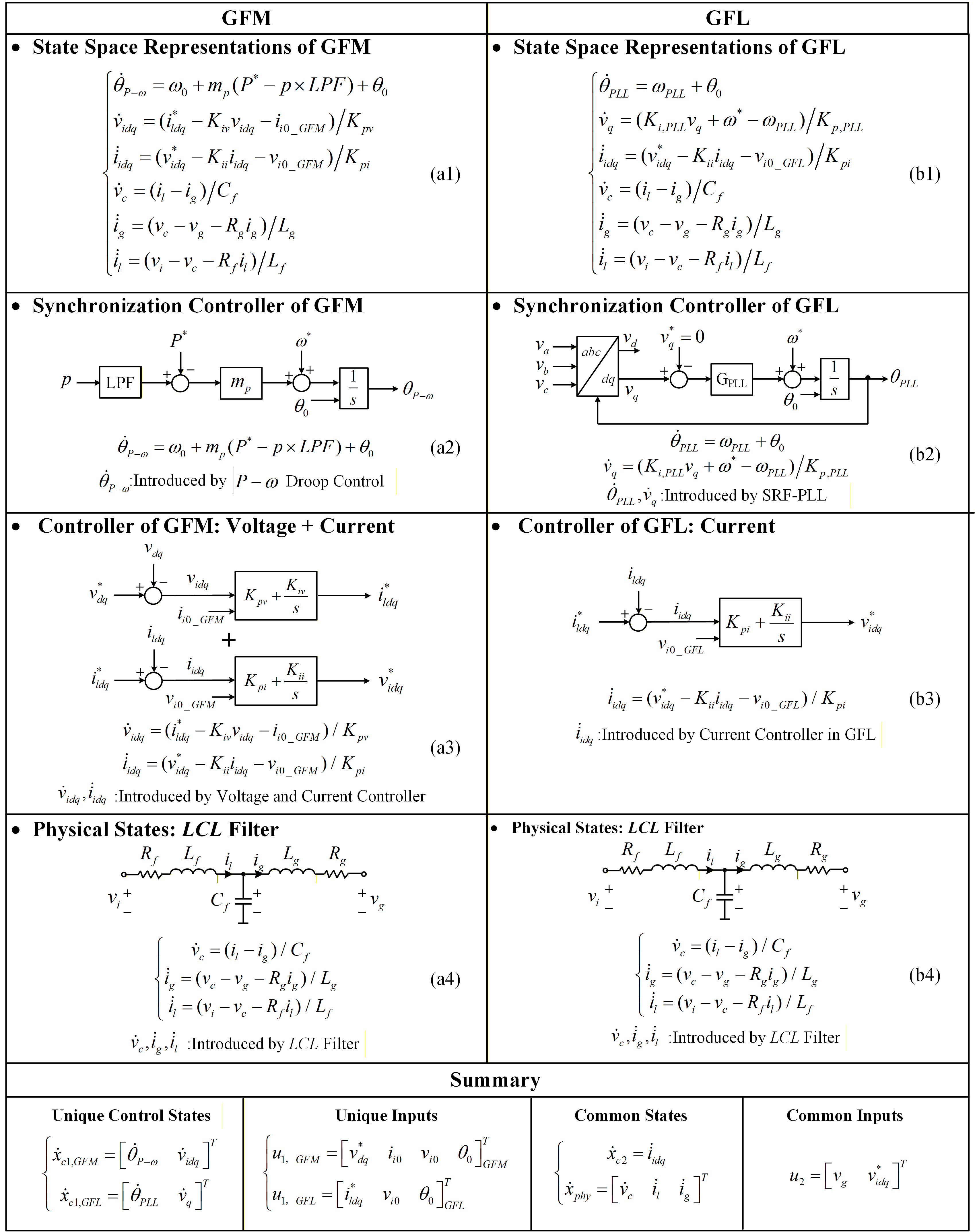}
\caption{Unified State Space Representations for GFL and GFM.}
\label{Fig:state-space_representations}
\end{figure*}

In (\ref{AB}) and (\ref{xc1}), the states $x=[x_{c1} \quad x_{c2} \quad x_{phy}]^T$ are divided into three parts: the independent state $x_{c1}$, the common state $x_{c2}$, and the physical state $x_{phy}$. The independent state $x_{c1}$ represents the unique control state variables of the system, such as the $P-\omega$ droop controller and voltage loop in GFM, and PLL in GFL. Thus, the state $x_{c1,GFM}$ can be expressed as consisting of the $P-\omega$ droop controller $\theta_{P-\omega}$ and the voltage loop state $v_{idq}$, while the state $x_{c1,GFL}$ consists of the PLL states $\theta_{PLL}$, and $v_q$. The common state $x_{c2}$ represents the common state of the system, such as the common current loop state $i_{idq}$, as shown in \figref{Fig:totalworkl}. The physical state $x_{phy}$ represents the state of the common $LCL$ filter of the system, consisting of the capacitor voltage $v_c$, grid-side inductor current $i_g$, and inverter-side inductor current $i_l$. Similarly, the system inputs are divided into two parts in (\ref{AB}): independent inputs and common inputs. The common inputs consist of the grid-side voltage $v_g$ and the inverter output $v_i$, while the independent inputs mainly consist of the initial values of integrators and the reference inputs in GFL and GFM. \figref{Fig:state-space_representations} provides a detailed illustration of the control architecture and state-space model construction for GFL and GFM, and offers a comprehensive summary of the aforementioned independent states, common states, physical states, common inputs, and independent inputs.

\begin{figure*}[th!]
\centering
\includegraphics[scale=0.385]{figure/eqpoint_high.png}
\caption{Derivation of the Equilibrium Points and the Domain of Attraction for GFL and GFM.}
\label{Fig:eqpoint}
\end{figure*}

With the summary in \figref{Fig:state-space_representations}, the analysis of smooth transitions between GFL and GFM has more specific objectives, which provides a research foundation for the following sections.

\subsection{Modeling Mode Transition}\label{Stability Analysis}

According to Section \ref{Modeling Mode Transition}, the detailed state-space representations for GFL and GFM have been obtained. Based on these representations, this section will adopt Lyapunov's theorems to analyze the equilibrium points and the domain of attraction of GFL and GFM, respectively. Thus, the analysis of smooth transitions between GFL and GFM can be reduced to an analysis of the mapping relationship between their equilibrium points and the relative positional relationship between the equilibrium points and their domain of attraction.

In order to obtain the equilibrium points of GFL and GFM, it is necessary to perform a stability analysis on GFL and GFM. A factor that plays a vital role in the stability of the system is the grid synchronization unit. For the grid-synchronization of grid-connected inverters, various synchronization methods are proposed. In this study, the synchronous method in GFL is considered as SRF-PLL, and the method in GFM is considered as $P-\omega$ droop controller. The block diagrams of the SRF-PLL and the $P-\omega$ droop controller are shown in \figref{Fig:state-space_representations}.The detailed derivation process is shown in \figref{Fig:eqpoint}.

\begin{figure}[th!]
\centering
\includegraphics[scale=0.6]{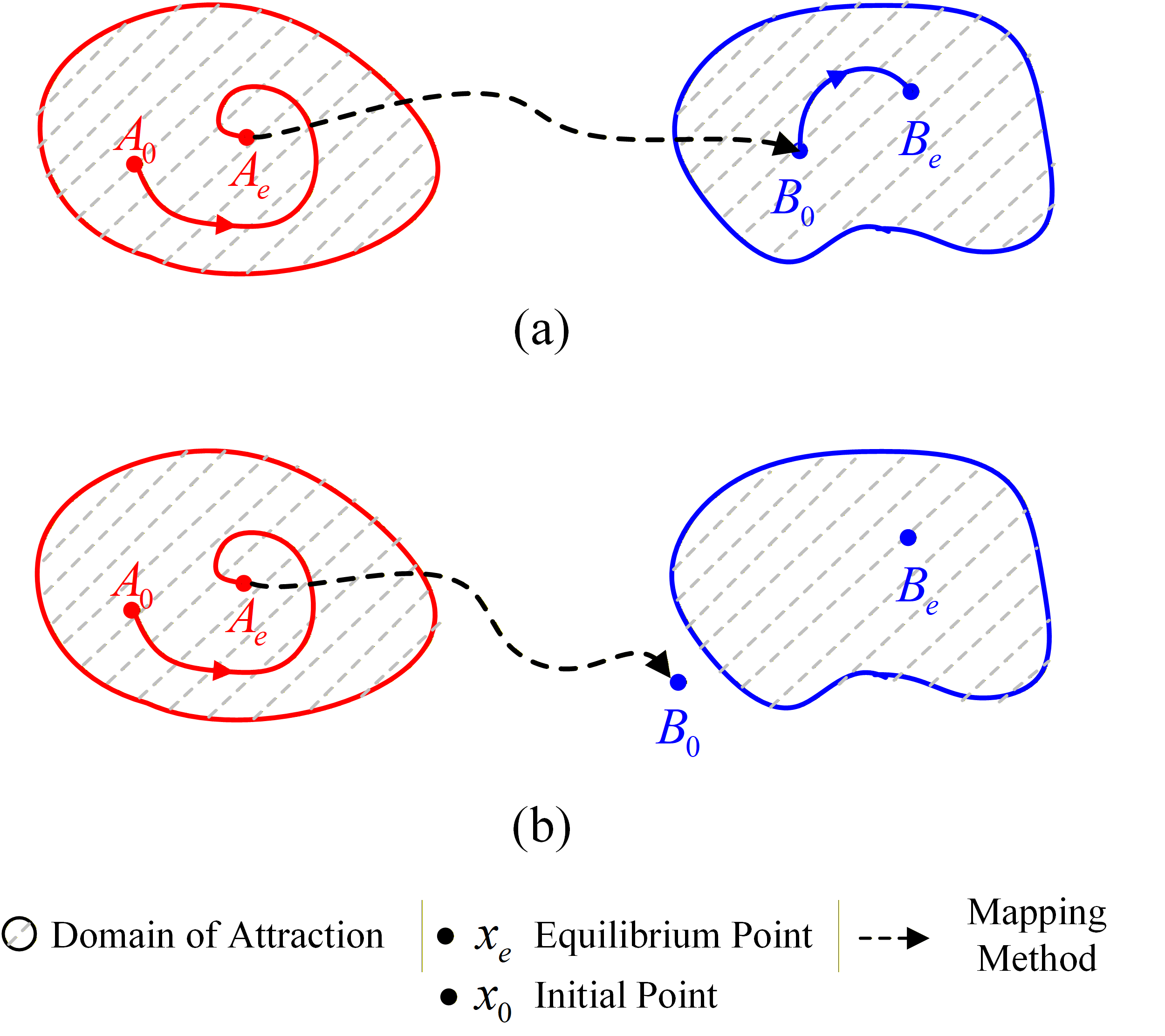}
\caption{The Essence of Mode Transitions.}
\label{Fig:nomapping}
\end{figure}

Based on the above derivations in \figref{Fig:eqpoint}, the information on the equilibrium points and domain of attraction of the system before and after mode transition is obtained. Accordingly, the modeling of mode transition is equivalent to investigating the mapping method for equilibrium points and the stability of the system after using the method, as shown in \figref{Fig:nomapping}.

In \figref{Fig:nomapping}, the transition between GFL and GFM is equivalent to the mapping from $A_e$ to $B_0$ (indicated by the dashed line in the figure), which illustrates two mapping results. In \figref{Fig:nomapping}(a), $A_e$ reaches $B_0$ through the mapping method (mode transition), and $B_0$ falls within the domain of attraction of Mode B. Then, the system will undergo a transient state to reach the steady state of Mode B, corresponding to the process from $B_0$ to $B_e$ in \figref{Fig:nomapping}(a), while the transient state is determined by the mapping method (mode transition method). In \figref{Fig:nomapping}(b), since $B_0$ falls outside the domain of attraction, the system will not be stable after undergoing equilibrium point mapping (mode transition).

\subsection{State Mapping Method}\label{EMM}

\begin{figure}[th!]
\centering
\includegraphics[scale=0.6]{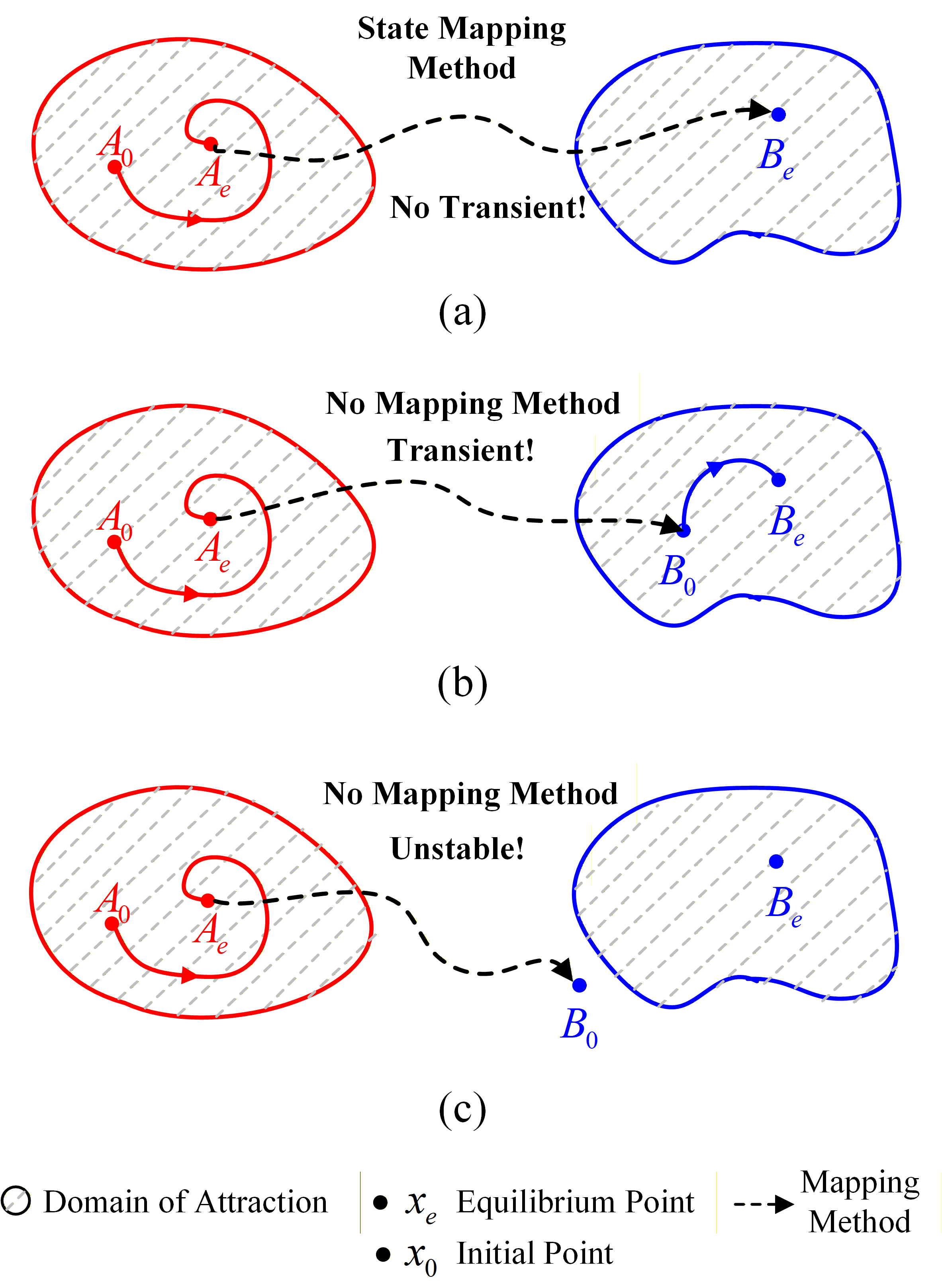}
\caption{Comparison of Whether to Apply State Mapping Method.}
\label{Fig:mapping}
\end{figure}

Based on Section \ref{Modeling Mode Transition} and Section \ref{Stability Analysis}, this section proposes a mapping method called \textit{state mapping method}, which achieves smooth transitions between GFM and GFL through the systematic approach, as shown in \figref{Fig:mapping}.

By applying the \textit{state mapping method}, the equilibrium point of the system before transition is directly mapped to the equilibrium point of the system after transition, as shown by $A_e$ to $B_e$ in \figref{Fig:mapping}(a). Since the system directly enters the steady state after the transition, there will be no transient! How to apply the \textit{state mapping method} will be introduced in detail.

\begin{figure*}[th!]
\centering
\includegraphics[scale=0.45]{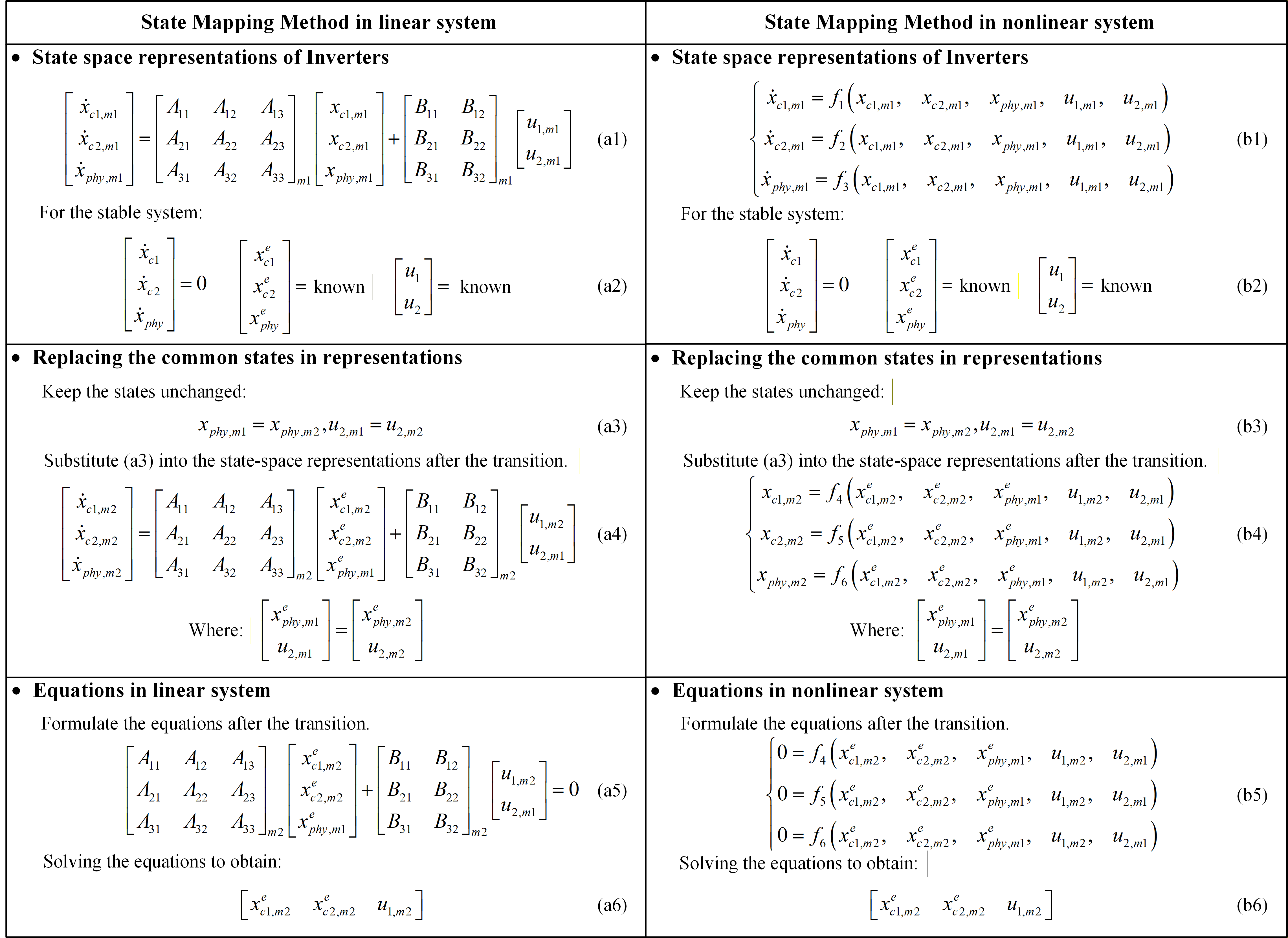}
\caption{Comparison of the Application of \textit{state mapping method} in Linear and Nonlinear Systems.}
\label{Fig:statemappingmethod}
\end{figure*}

In Section \ref{Modeling Mode Transition}, the representations for the unified state-space modeling of the system have been provided, in which the state variables $x$ and the inputs $u$ are summarized in detail. In order to achieve smooth transitions without transients, the focus should be on the changes in the common states and inputs before and after transitions. In \figref{Fig:state-space_representations}, the common states and inputs refer to $x_{phy}$ and $u_2$. Therefore, achieving smooth transitions between GFM and GFL requires ensuring that $x_{phy}$ and $u_2$ remain unchanged before and after the transition. \figref{Fig:statemappingmethod} shows the procedure for applying the \textit{state mapping method} in linear and nonlinear systems, respectively.

Here, taking the transition from GFL to GFM as an example to introduce the \textit{state mapping method}. Table \ref{tab1} shows the system parameters under GFL.

\begin{table}[!h]\label{tab1}
\caption{System Parameters of GFL}
\centering
\renewcommand{\arraystretch}{1.5} 
\setlength{\tabcolsep}{10pt} 
\small 
\begin{tabular}{|c||c||c|}
\hline
\multicolumn{3}{|c|}{\textbf{$LCL$ Parameters}}\\
\hline
$v^e_{cdq}$ & Capacitor Voltage & 0.92/0 p.u. \\
\hline
$i^e_{ldq}$ & Inverter-side Current & 0.51/0.02 p.u. \\
\hline
$i^e_{gdq}$ & Grid-side Current & 0.51/0 p.u. \\
\hline
\multicolumn{3}{|c|}{\textbf{Unique Inputs}}\\
\hline
$i^*_{ldq}$ & Current Reference & 0.55/0 p.u. \\
\hline
$v_{i0dq}$ & Initial of PI in Current Loop & 0/0 p.u. \\
\hline
$\theta_0$ & Initial Phase & 0 rad/s \\
\hline
\multicolumn{3}{|c|}{\textbf{Unique States}}\\
\hline
$\theta^e_{PLL}$ & Steady Phase of PLL & $\int \omega^* dt$ \\
\hline
$v^e_q$ & Q-axis Voltage & 0 p.u. \\
\hline
\multicolumn{3}{|c|}{\textbf{Common Inputs}}\\
\hline
$v_g$ & Grid Voltage & 1 p.u. \\
\hline
$v^*_{idq}$ & Output Voltage Reference & 0.93/0.09 p.u. \\
\hline
\end{tabular}
\end{table}

Ensuring that $x_{phy}$ and $u_2$ remain unchanged, the obtained GFM parameters are shown in Table \ref{tab2}.

\begin{table}[!h]\label{tab2}
\caption{Applying State Mapping Method}
\centering
\renewcommand{\arraystretch}{1.5} 
\setlength{\tabcolsep}{10pt} 
\small 
\begin{tabular}{|c||c||c|}
\hline
\multicolumn{3}{|c|}{\textbf{Unique States}}\\
\hline
$\theta^e_{P-\omega}$ & Steady Phase of $P-\omega$ & $\int \omega^* dt$ \\
\hline
$v^e_{idq}$ & Voltage Steady-State Error & 0/0 p.u. \\
\hline
\multicolumn{3}{|c|}{\textbf{Unique Inputs}}\\
\hline
$v^*_{dq}$ & Voltage Reference & 0.92/0 p.u. \\
\hline
$v_{i0dq}$ & Initial of PI in Current Loop & 0.94/0.02 p.u. \\
\hline
$i_{i0dq}$ & Initial of PI in Voltage Loop & 0.92/0.094 p.u. \\
\hline
$\theta_0$ & Initial Phase & -0.226 rad/s \\
\hline
\end{tabular}
\end{table}

\begin{figure}[!h]
\centering
\includegraphics[scale=0.5]{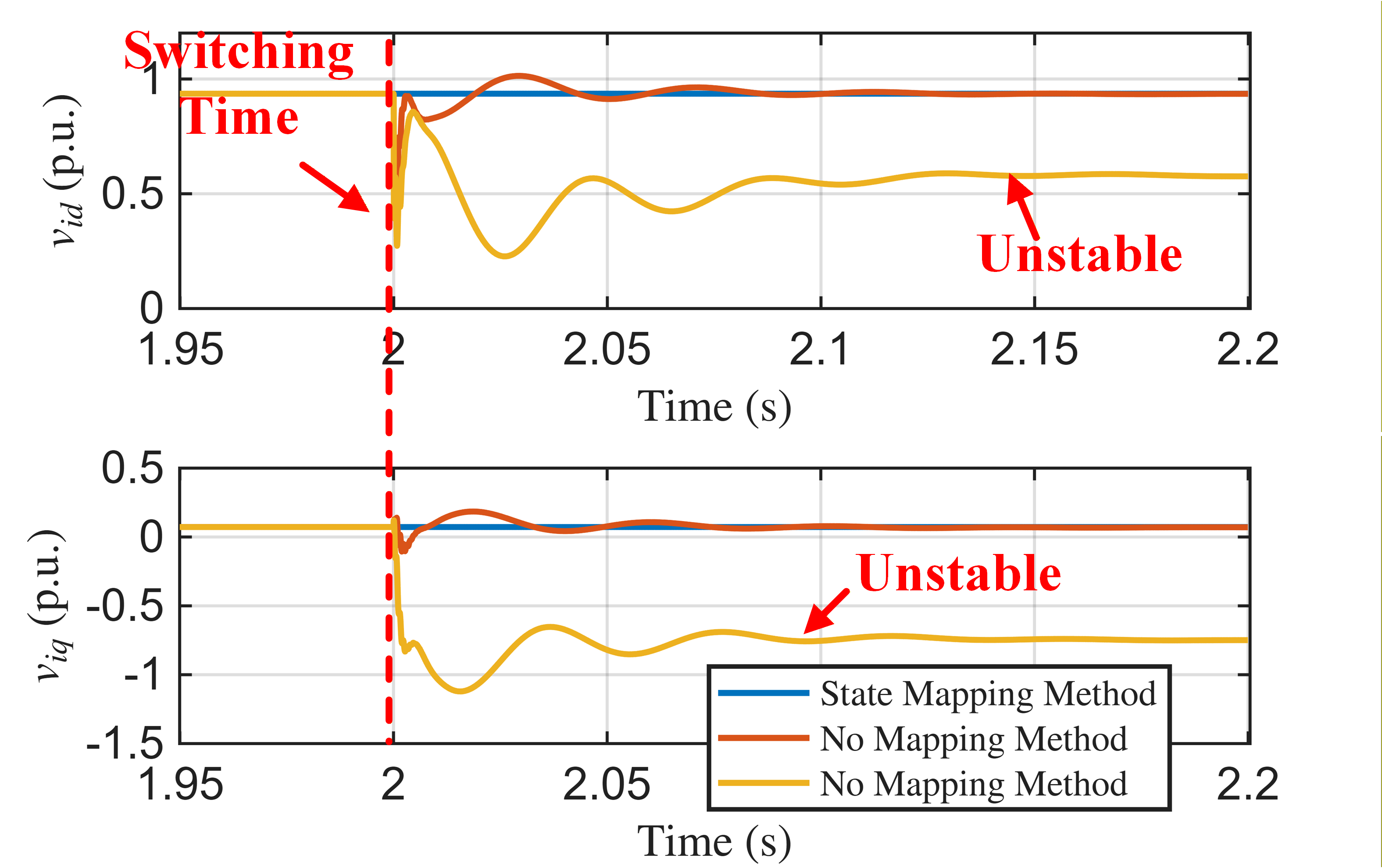}
\caption{Comparison of Output Voltage With and Without Applying State Mapping Method.}
\label{Fig:Comparevidq}
\end{figure}

Through the \textit{state mapping method}, the conditions that satisfy the smooth transition of the system from GFL to GFM are obtained. These conditions include the voltage reference value, the initial values of the integrators in both the voltage loop and the current loop, and the initial phase value in the synchronous controller. \figref{Fig:Comparevidq} shows the process of the output voltage undergoing the transition from GFL to GFM with or without using the \textit{state mapping method}, corresponding to the three scenarios in \figref{Fig:mapping}.

As illustrated in \figref{Fig:Comparevidq}, in contrast to the other two scenarios, the system exhibits virtually no transient response following the transition from GFL to GFM when the State Mapping Method is employed. The scenario from GFM to GFL will not be elaborated further due to similar principles, and the detailed results are presented in Section \ref{Section:Experimental Results}.

\section{Control Design for Smooth Mode Transition}\label{Control Design for Smooth Mode Transition}

Based on the \textit{state mapping method} in Section \ref{Analysis of Mode Transition Transients}, it is clearly demonstrated that to achieve a smooth mode transition, attention should be focused on two types of parameters: phase and amplitude. Thus, the controllers are divided into the synchronization controller and the amplitude controller to achieve a smooth transition of the system. In this section, the synchronization controller and the amplitude controller will be introduced separately. In order to elucidate the distinction between the synchronization controller and the amplitude controller, this section also solely examines the impact of synchronization control on mode transitions, while amplitude control is consistently maintained.

\subsection{Synchronization Control}\label{Synchronization Control}

The synchronization loop for both grid-following and grid-forming control is illustrated in \figref{Fig:totalworkl}, as shown in (\ref{theta}). 

\begin{equation}\label{theta}
\begin{cases}
\omega_{PLL} = \omega^* + (K_{p,PLL}v_q + K_{i,PLL}\int v_q dt) \\[6pt]
\omega_{P-\omega} = \omega^* + m_p(P^*-p\times LPF) \\
\end{cases}
\end{equation}

In order to achieve phase synchronization via the synchronization controller, the most straightforward method is to adjust the initial value of the integrator $\theta_0$, as shown in (\ref{theta0}), thereby achieving phase synchronization during the mode transition.

\begin{equation}\label{theta0}
\theta = \int \omega dt + \theta_{0} 
\end{equation}

\begin{figure}[th!]
\centering
\includegraphics[scale=0.65]{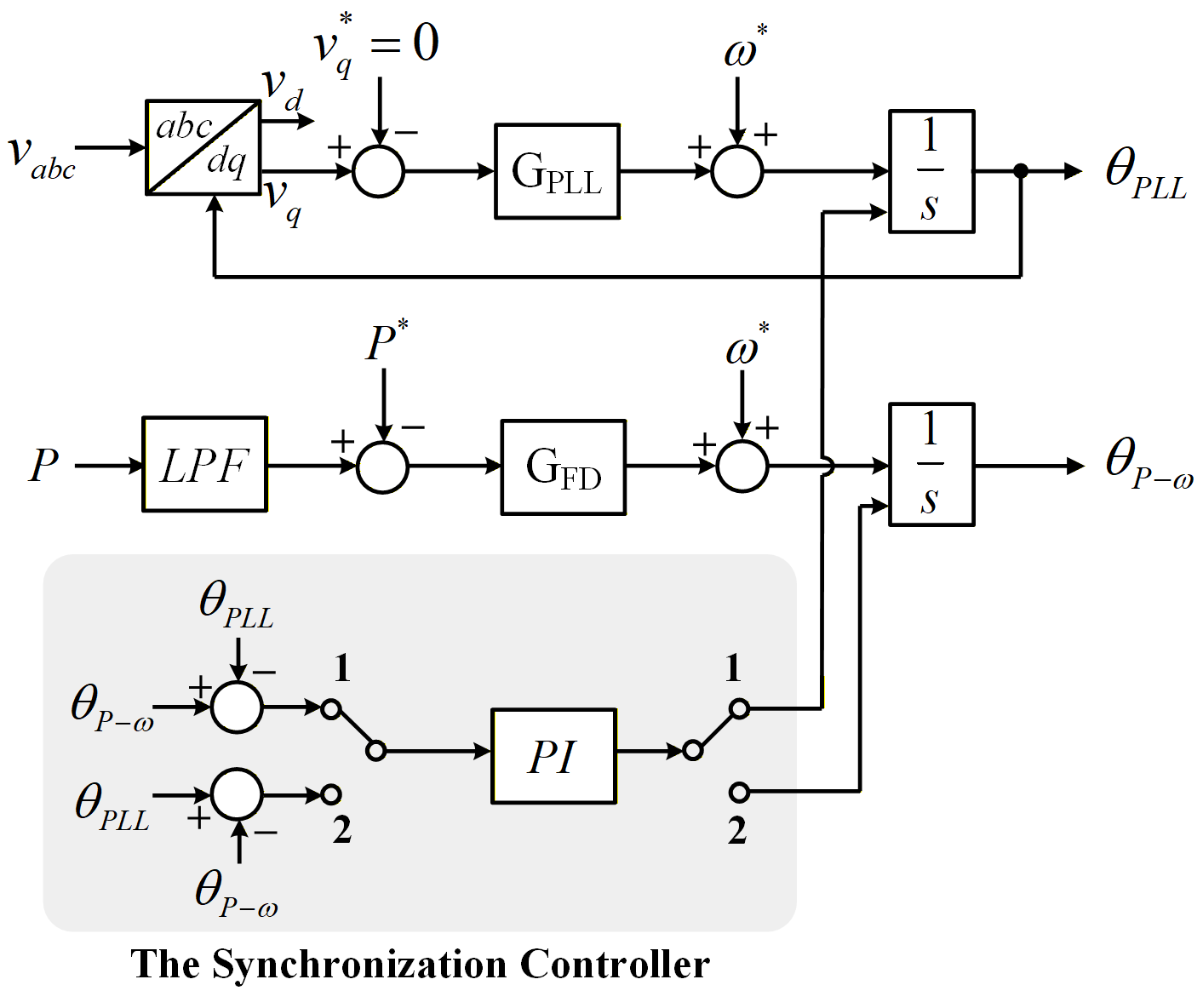}
\caption{Controller for synchronization control.}
\label{Fig:controller}
\end{figure}

In order to achieve control over the initial phase value in the synchronization loop, a dedicated controller is designed that enables the system to measure the phase difference and achieve compensation by changing the initial value at the moment of transition, as shown in \figref{Fig:controller}. When switching from grid-forming to grid-following control, select Position 1. Conversely, when switching from grid-following to grid-forming control, select Position 2.

\begin{figure}[th!]
\centering
\includegraphics[scale=0.55]{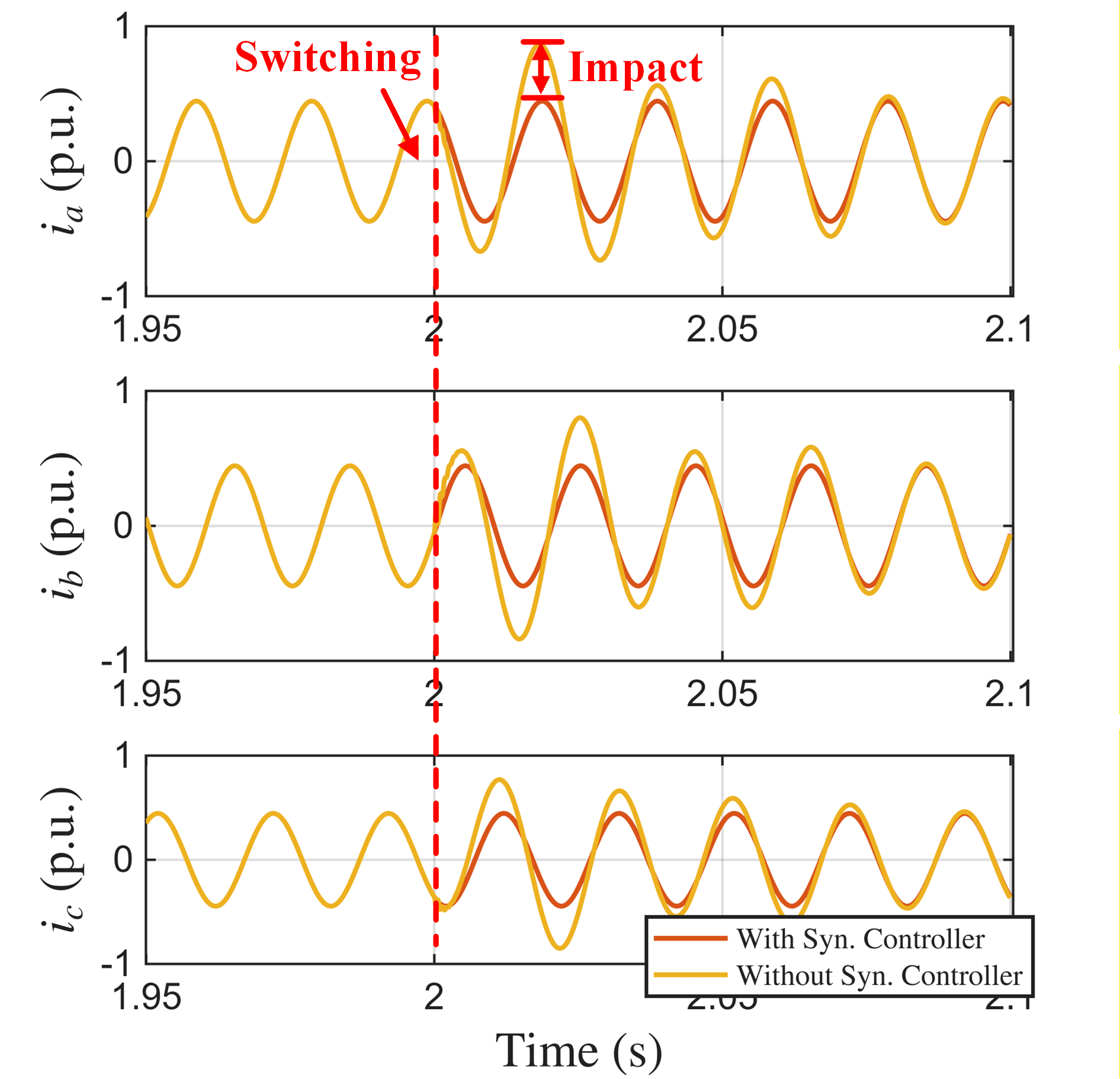}
\caption{Comparison of current variations from grid-following to grid-forming control.}
\label{Fig:SynControl_GFLtoGFM}
\end{figure}

\begin{figure}[th!]
\centering
\includegraphics[scale=0.55]{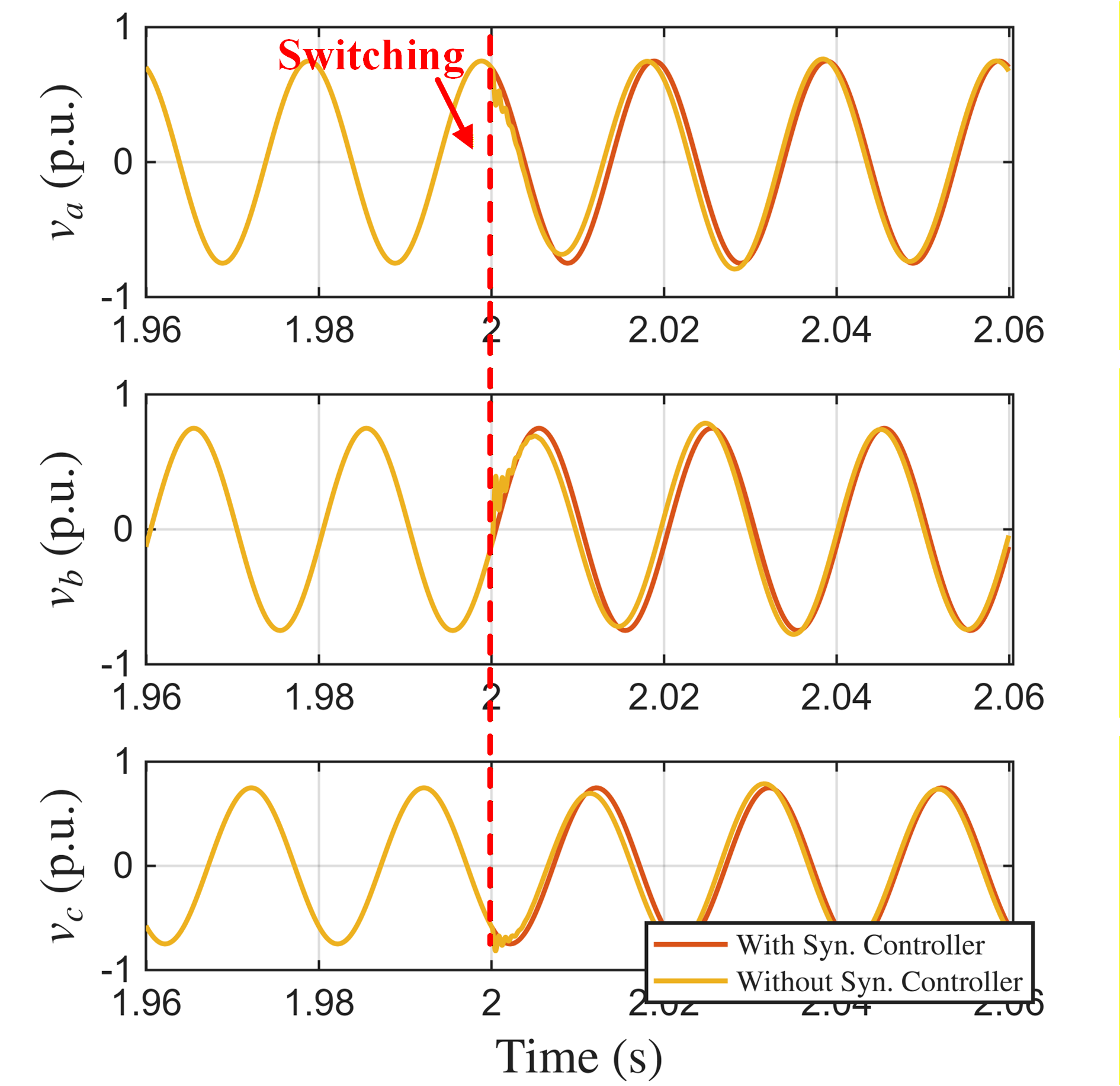}
\caption{Comparison of voltage variations from grid-following to grid-forming control.}
\label{Fig:SynControl_GFLtoGFM_va}
\end{figure}

By applying the designed controller, \figref{Fig:SynControl_GFLtoGFM} and \figref{Fig:SynControl_GFLtoGFM_va} present the resulting changes in current and voltage in the $dq$ frame during the transition from GFL to GFM, with the amplitude controller included. It is worth noting that during the transition from GFM to GFL, since the characteristic of $P-\omega$ droop controller is to autonomously control the system frequency, while the working principle of the phase-locked loop (PLL) is to follow the system frequency, it results in the phase of GFL always being locked to the phase of GFM. Consequently, when the amplitude controller is used, the synchronization controller plays a limited role in the smooth transition from GFM to GFL.

As shown in Figures \ref{Fig:SynControl_GFLtoGFM} and \ref{Fig:SynControl_GFLtoGFM_va}, with the amplitude controller already in place, the addition of the synchronization controller significantly reduces transients during the mode transition, except in the case of switching from GFM to GFL, where its effect is limited. The specific experiments are detailed in Section \ref{Section:Experimental Results}.

\subsection{Amplitude Control}

In contrast to Section \ref{Synchronization Control}, this section focuses on the impact of the amplitude controller on the mode transition, which implies that the discussion in this section includes the synchronization controller.

In the absence of the amplitude controller, the conventional PI controller can be expressed by (\ref{y}). Due to the differences in control loop configurations between GFL and GFM, the amplitude of the controller output inevitably exhibits a sudden change after transition, which consequently affects the smoothness of the mode transition. To mitigate the impact caused by amplitude jumps, the initial value of the integrator is selected as the control subject for the amplitude controller based on (\ref{y}).

\begin{equation}\label{y}
y = k_p(x^*-x)+k_i\int(x^*-x)dt+x_0
\end{equation}

\begin{figure}[th!]
\centering
\includegraphics[scale=0.45]{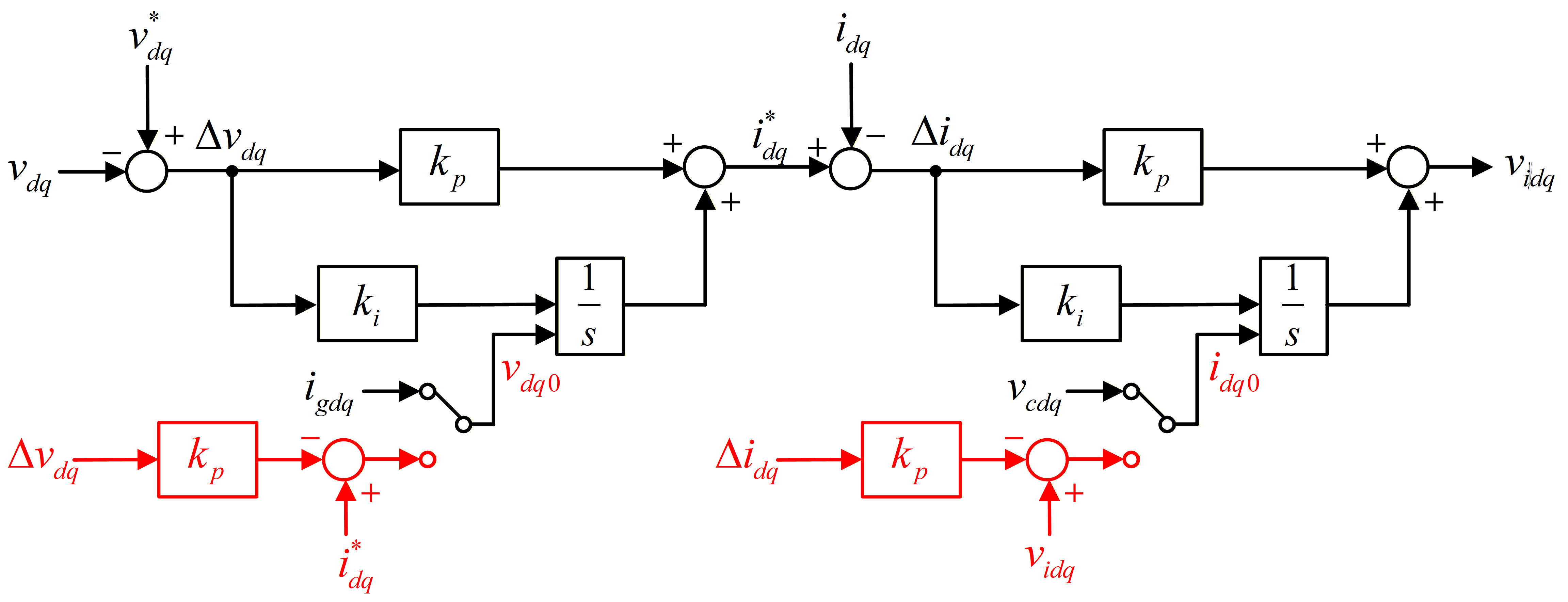}
\caption{The improved PI controller.}
\label{Fig:Improved_PI_Controller}
\end{figure}

To prevent sudden changes in the controllers' outputs, a modified PI controller is designed to achieve this control objective, as shown in \figref{Fig:Improved_PI_Controller}.

In \figref{Fig:Improved_PI_Controller}, the portion responsible for the amplitude controller is highlighted in red. Its function is to use the difference between the pre-steady‑state output of the controller and the real‑time error as the initial value of the integrator for the subsequent control mode during mode transition, as indicated by (\ref{x_0}).

\begin{equation}\label{x_0}
x_0 = y - k_p(x^*-x)
\end{equation}

\begin{figure}[th!]
\centering
\includegraphics[scale=0.5]{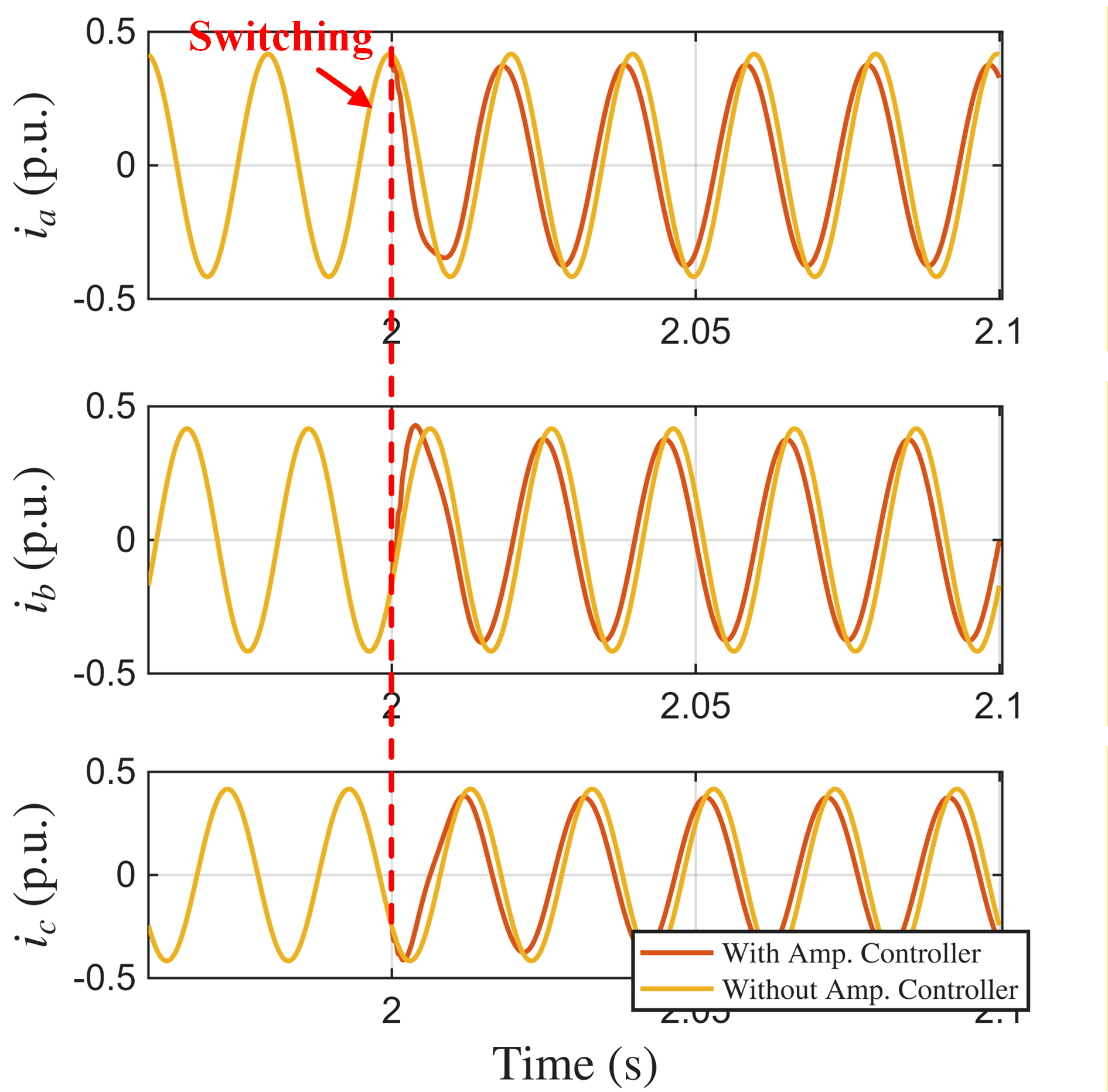}
\caption{Comparison of current variations from grid-forming to grid-following control.}
\label{Fig:GFMtoGFL_ia}
\end{figure}

\begin{figure}[th!]
\centering
\includegraphics[scale=0.5]{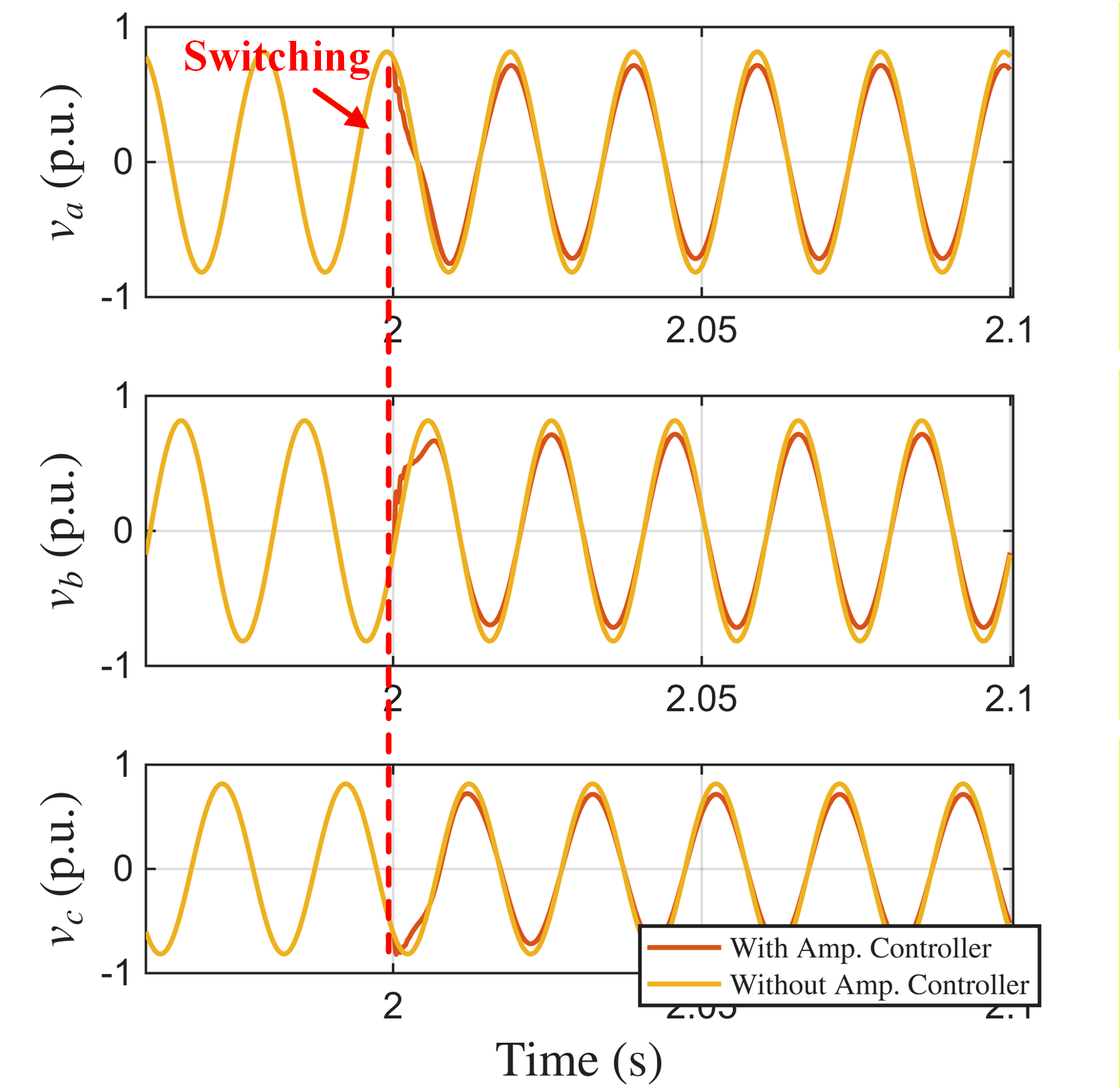}
\caption{Comparison of voltage variations from grid-forming to grid-following control.}
\label{Fig:GFMtoGFL_va}
\end{figure}

\begin{figure}[th!]
\centering
\includegraphics[scale=0.5]{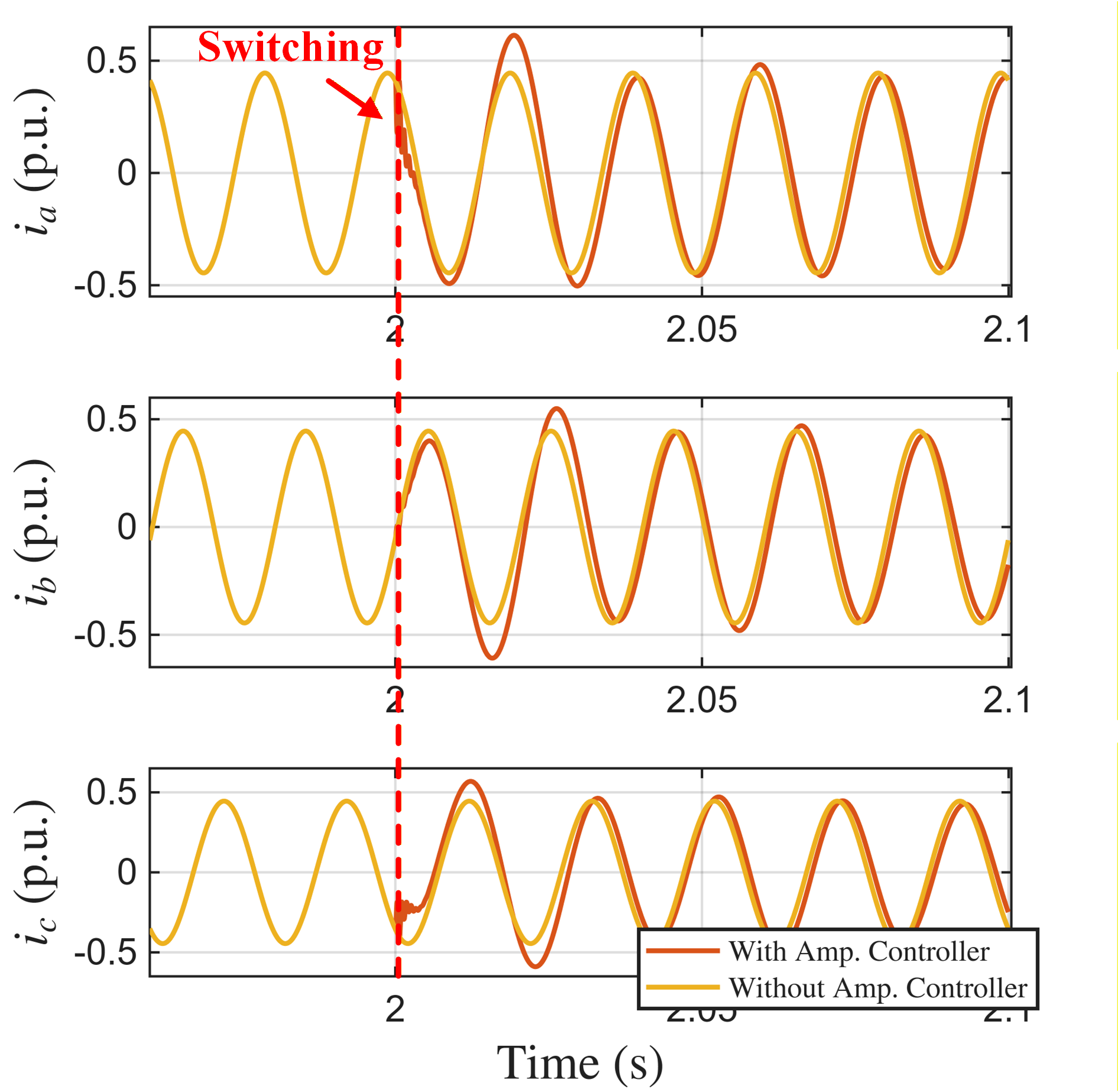}
\caption{Comparison of current variations from grid-following to grid-forming control.}
\label{Fig:GFLtoGFM_ia}
\end{figure}

\begin{figure}[th!]
\centering
\includegraphics[scale=0.5]{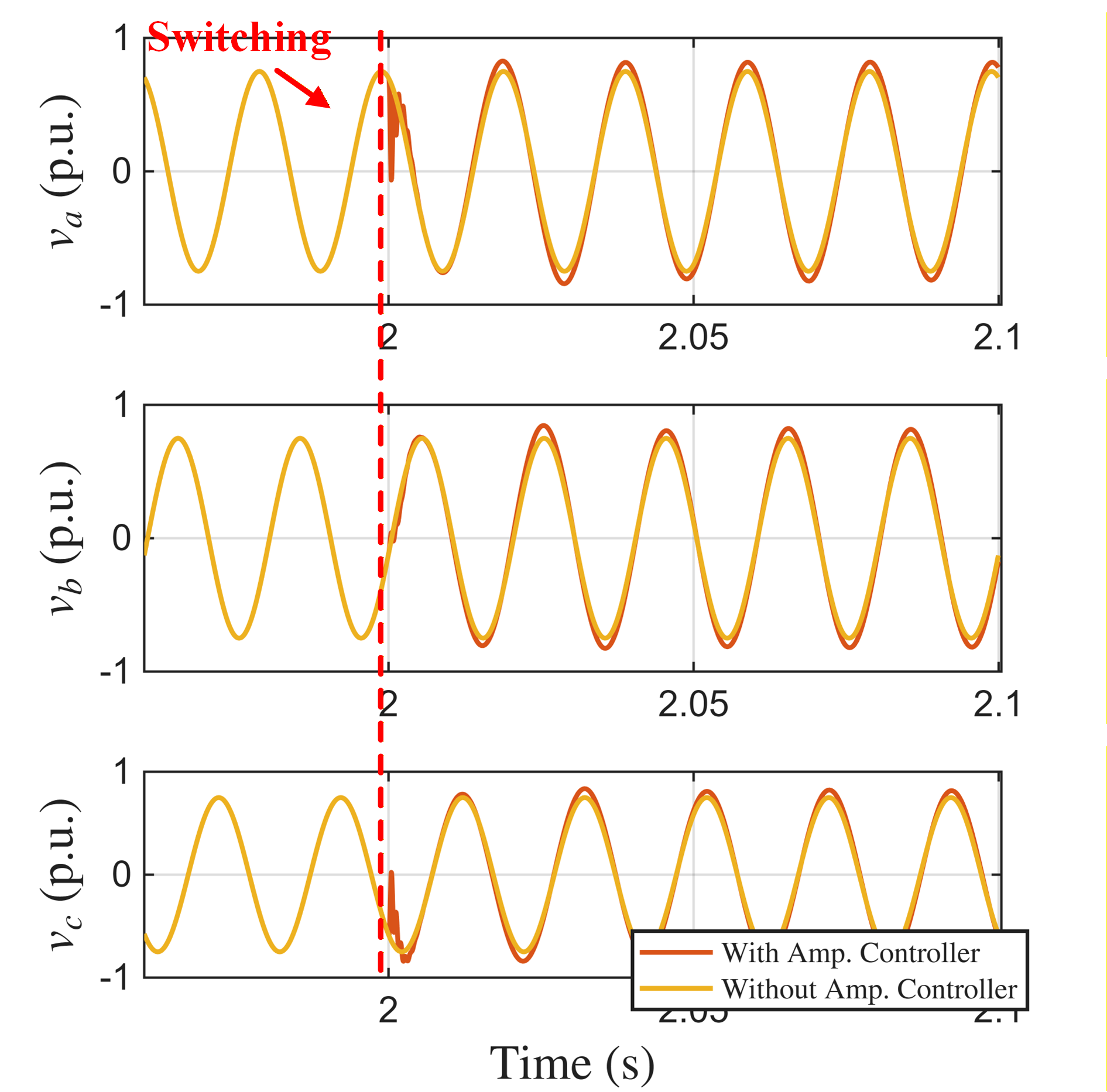}
\caption{Comparison of voltage variations from grid-following to grid-forming control.}
\label{Fig:GFLtoGFM_va}
\end{figure}

\figref{Fig:GFMtoGFL_ia} and \figref{Fig:GFMtoGFL_va} present the resulting changes in current and voltage during the transition from GFM to GFL, with the amplitude controller excluded. Meanwhile, \figref{Fig:GFLtoGFM_ia} and \figref{Fig:GFLtoGFM_va} present the resulting changes from GFM to GFL.

Even with synchronization control, the application of amplitude control mitigates the abruptness of current and voltage changes during mode transitions. The specific experiments are detailed in Section \ref{Section:Experimental Results}.

\section{Experimental Results} \label{Section:Experimental Results}

\begin{figure}[th!]
\centering
\includegraphics[scale=0.25]{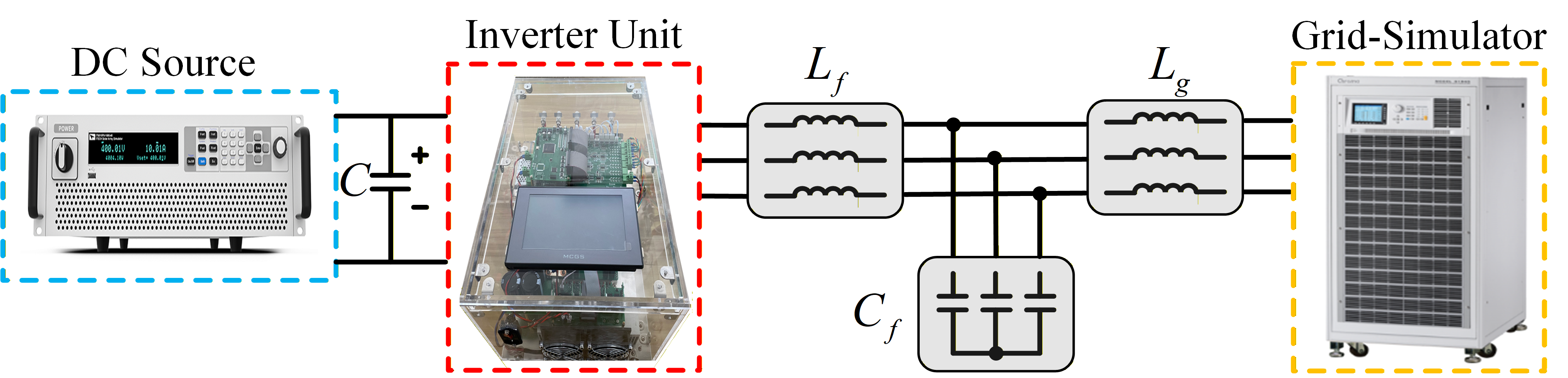}
\caption{Hardware platform of the single-inverter-infinite-bus-system.}
\label{Fig:setup}
\end{figure}

To verify the effectiveness of the proposed method, a three-phase grid-connected system, as shown in \figref{Fig:setup}, is employed for the experiments. The parameters of the experimental setup are shown in Table \ref{tab3}.

\begin{figure*}[th!]
\centering
\begin{minipage}{0.48\textwidth}
    \centering
    \includegraphics[width=\linewidth]{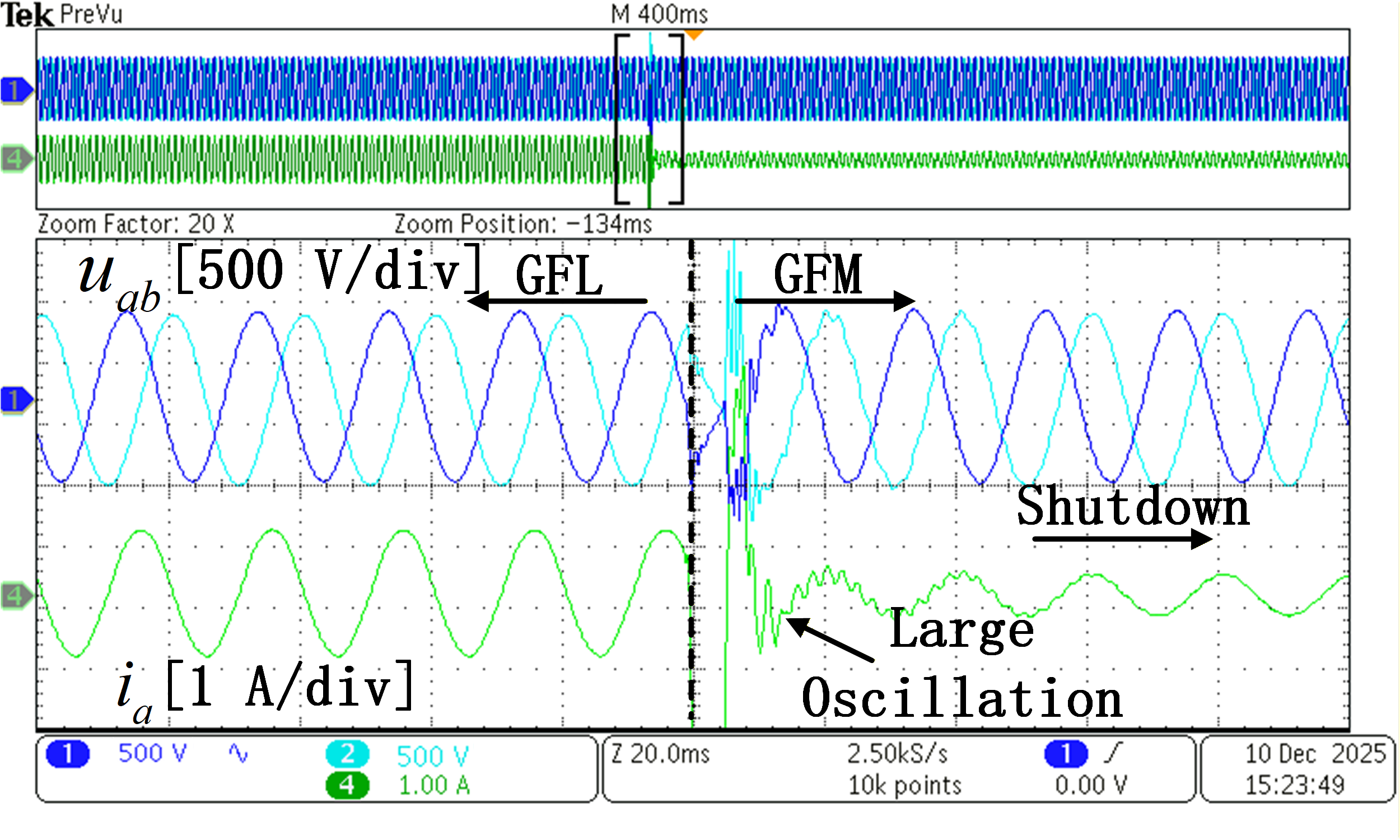}
    \caption{Mode transition from grid-following to grid-forming control without smooth method.}
    \label{Fig:GFLtoGFM_without_v4}
\end{minipage}
\hfill
\begin{minipage}{0.48\textwidth}
    \centering
    \includegraphics[width=\linewidth]{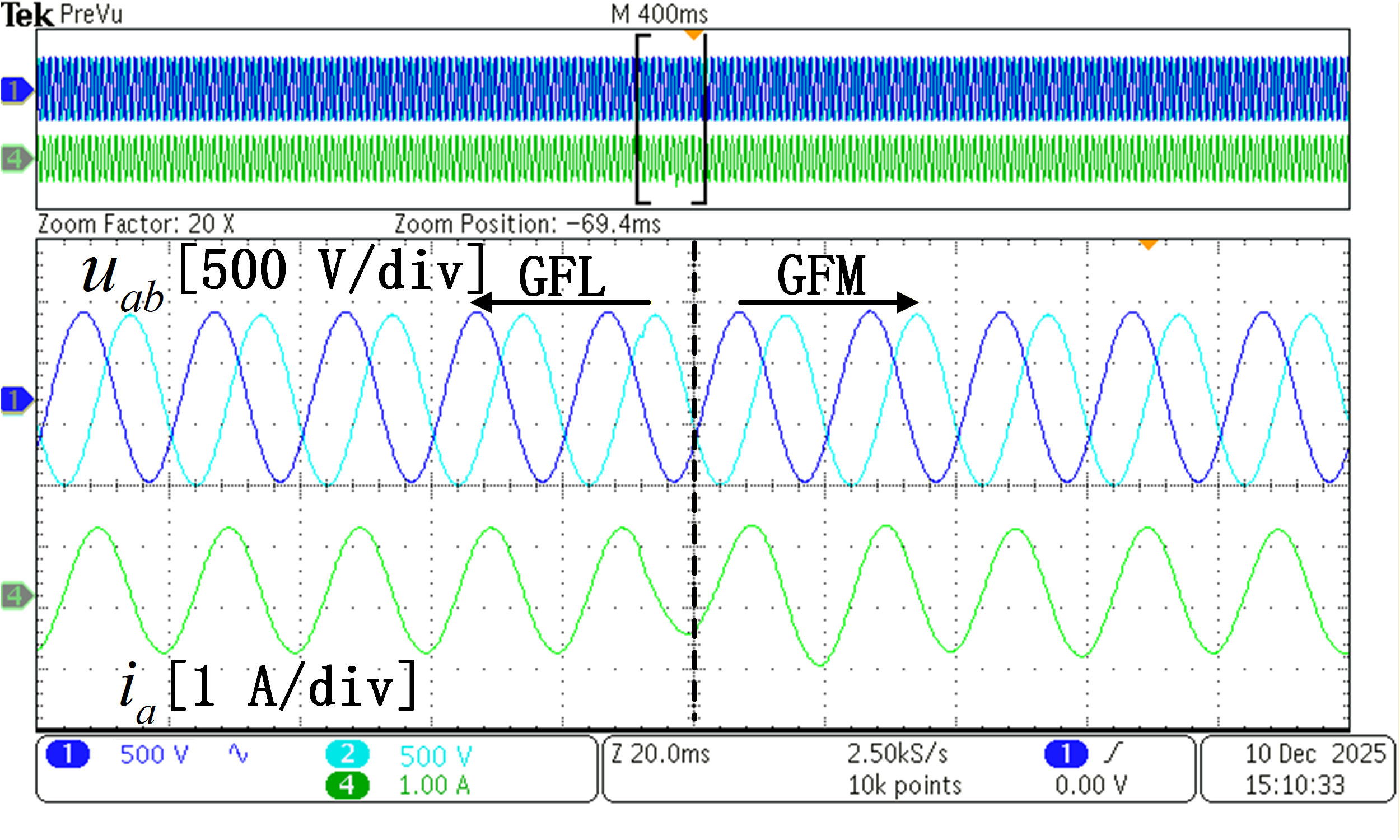}
    \caption{Smooth mode transition from grid-following to grid-forming control.}
    \label{Fig:GFLtoGFM_with_v4}
\end{minipage}
\end{figure*}

\begin{table}[t!]
\caption{Inverter System Parameters}
\centering
\renewcommand{\arraystretch}{1.5} 
\setlength{\tabcolsep}{12pt} 
\small 
\begin{tabular}{|c||c||c|}
\hline
\multicolumn{3}{|c|}{\textbf{AC Grid Parameters}}\\
\hline
$v_g$ & Grid Voltage & 200 V \\
\hline
$f_g$ & Grid Frequency & 50 Hz \\
\hline
\multicolumn{3}{|c|}{\textbf{DC Side Parameters}}\\
\hline
$C$ & DC Capacitor & 1500 $\mu$F \\
\hline
$v_{dc}$ & Constant DC Voltage & 450 V \\
\hline
$P$ & Nominal Power & 2 kW \\
\hline
\multicolumn{3}{|c|}{\textbf{LC Filter and Grid Line Impedance}}\\
\hline
$L_f$ & Filtering Inductor & 5 mH \\
\hline
$R_f$ & Inner Resistance of $L_f$ & 0.05 $\ohm$ \\
\hline
$C_f$ & Filtering Capacitor & 30 $\mu$F \\
\hline
$L_g$ & Line Inductor & 4 mH \\
\hline
$R_g$ &  Inner Resistance of $L_g$ & 0.4 $\ohm$ \\
\hline
\end{tabular}
\label{tab3}
\end{table}

\begin{figure*}[th!]
\centering
\begin{minipage}{0.48\textwidth}
    \centering
    \includegraphics[width=\linewidth]{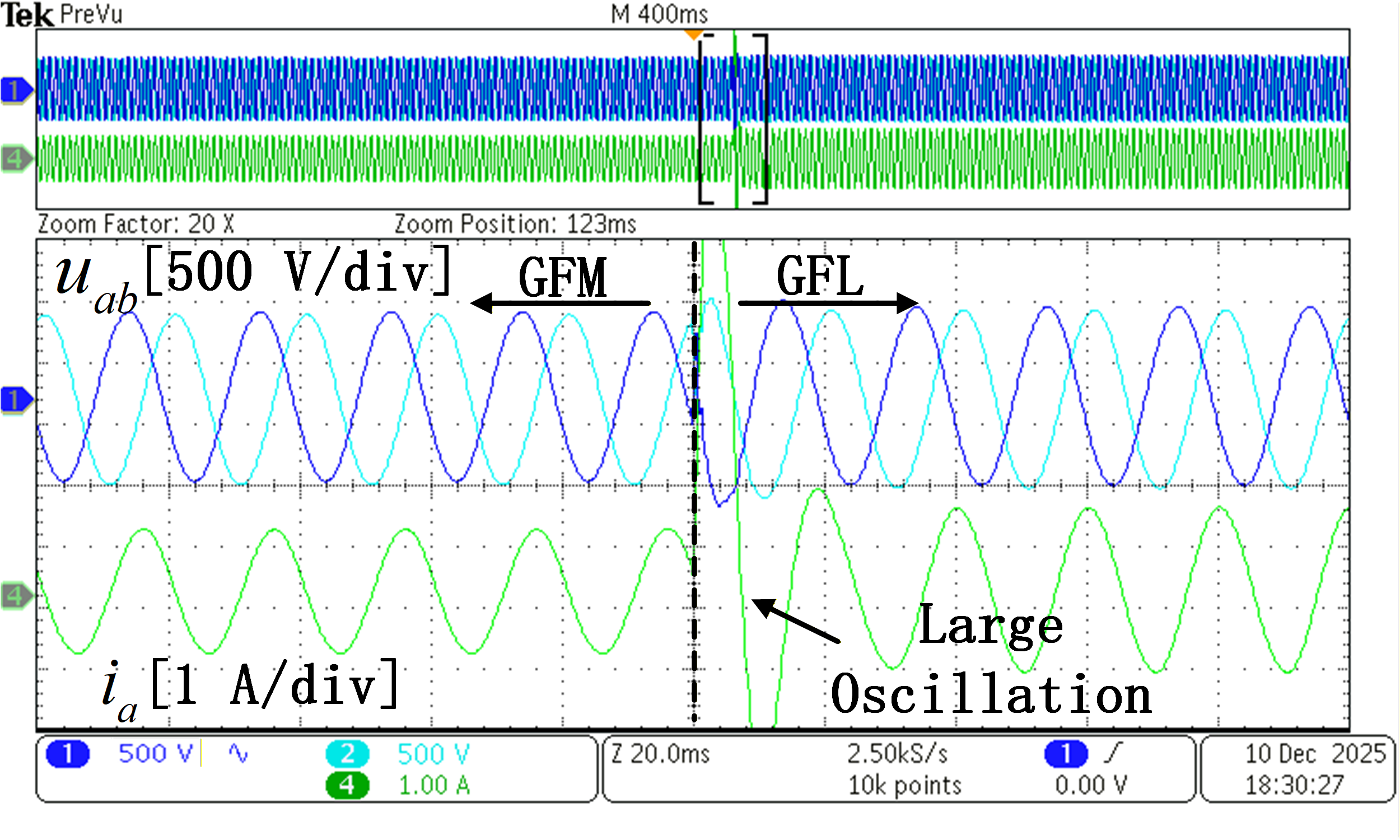}
    \caption{Mode transition from grid-forming to grid-following control without smooth method.}
    \label{Fig:GFMtoGFL_without_v4}
\end{minipage}
\hfill
\begin{minipage}{0.48\textwidth}
    \centering
    \includegraphics[width=\linewidth]{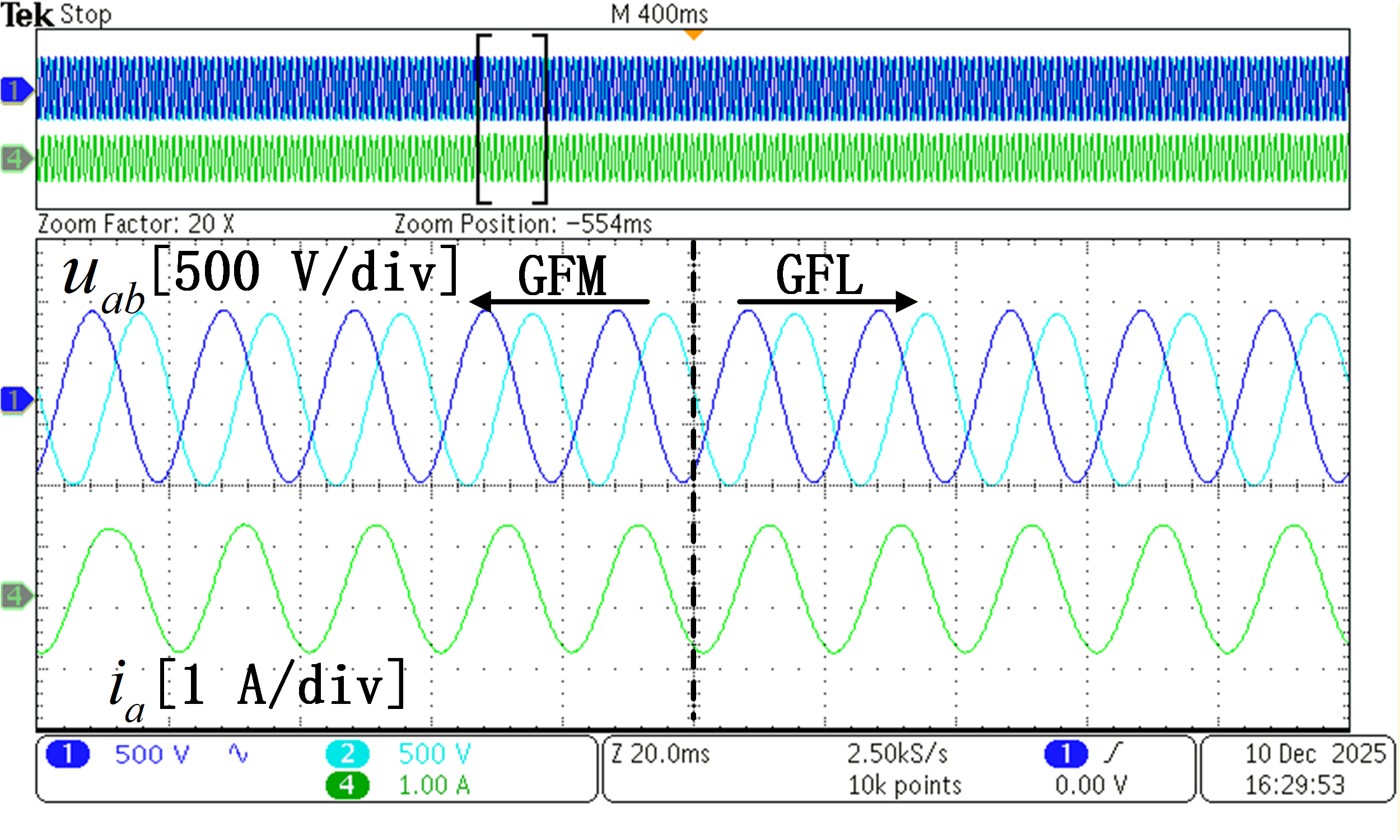}
    \caption{Smooth mode transition from grid-following to grid-forming control.}
    \label{Fig:GFMtoGFL_with_v4}
\end{minipage}
\end{figure*}

\subsection{Case 1: Experiments from Grid-Following to Grid-Forming} \label{Section:CASE1}

The experimental results of the transition from Grid-Following to Grid-Forming are as shown in \figref{Fig:GFLtoGFM_without_v4}-\figref{Fig:GFLtoGFM_with_v4}. During the transition, the experimental waveforms of the line voltages $U_{ab}$ and $U_{bc}$ and the converter current $i_a$ without the smooth mode transition control are shown in \figref{Fig:GFLtoGFM_without_v4}. During the transition, the converter outputs have large oscillations that can even lead to a shutdown of the converter. When the smooth mode transition method is applied, the experimental waveforms of the voltages and the converter current are shown in \figref{Fig:GFLtoGFM_with_v4}. In this case, both the PCC voltages and converter currents have realized a smooth mode transition. It is worth mentioning that the system operates at the same operating point before and after the transition.

\subsection{Case 2: Experiments from Grid-Forming to Grid-Following} \label{Section:CASE2}

During the transition from grid-forming to grid-following control, the experimental waveforms of PCC voltages and converter currents without the smooth mode transition are shown in \figref{Fig:GFMtoGFL_without_v4}. The experimental waveforms of PCC voltages and converter currents with the smooth mode transition method are shown in \figref{Fig:GFMtoGFL_with_v4}. Without the proposed method, the large oscillations during the transition leads to the trigger of the hardware protection. With the proposed control, it is clear that the proposed smooth mode transition method works well and gives obviously a much smoother transition between grid-forming control and grid-following control.

\subsection{Case 3: Experiments on Different Operating Points} \label{Section:CASE3}

In Section \ref{Section:CASE1} and \ref{Section:CASE2}, it is worth mentioning that the system operates at the same steady-state operating point both before and after the transition. To account for the fact that practical systems may operate at different operating points after the transition from grid-following to grid-forming control, experiments are also conducted at different static operating points. To achieve a smooth mode transition, the transition between different operating points must be transformed into an equivalent transition at the same operating point.

\begin{figure}[!h]
\centering
\includegraphics[scale=0.37]{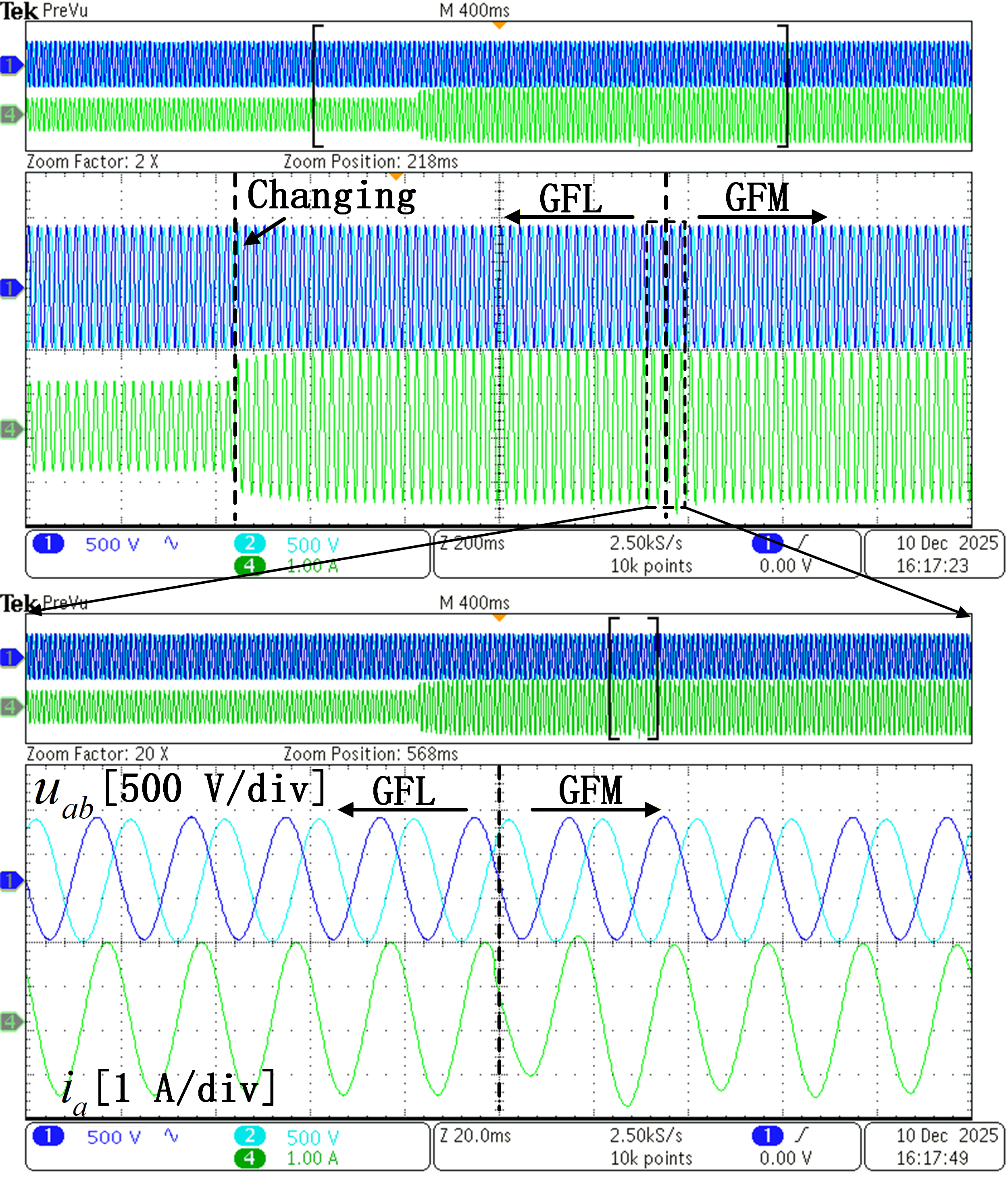}
\caption{Smooth mode transition from grid-following to grid-forming control}
\label{Fig:GFLtoGFM_after}
\end{figure}

\begin{figure}[!h]
\centering
\includegraphics[scale=0.37]{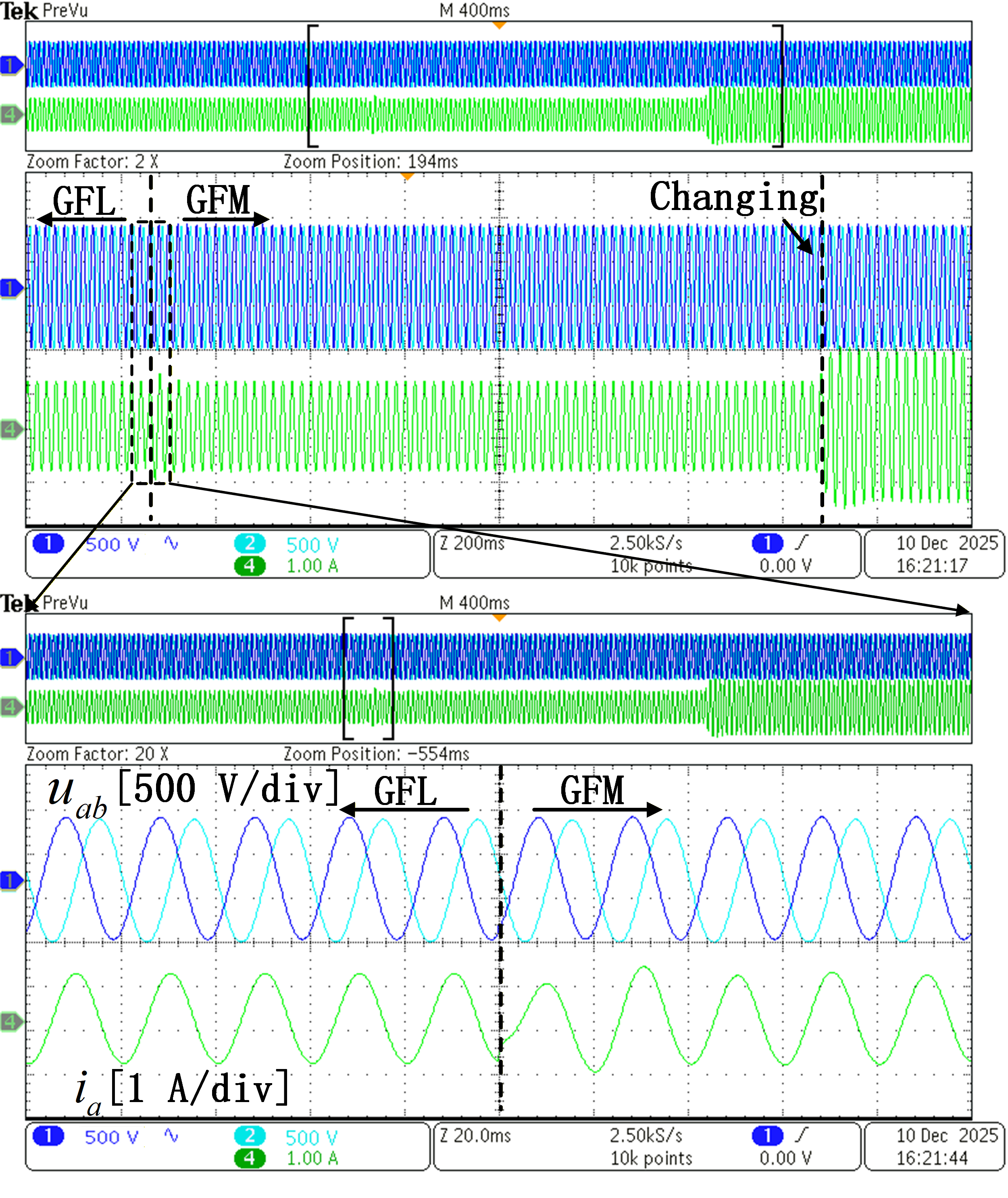}
\caption{Smooth mode transition from grid-following to grid-forming control}
\label{Fig:GFLtoGFM_before}
\end{figure}

\begin{figure}[th!]
\centering
\includegraphics[scale=0.37]{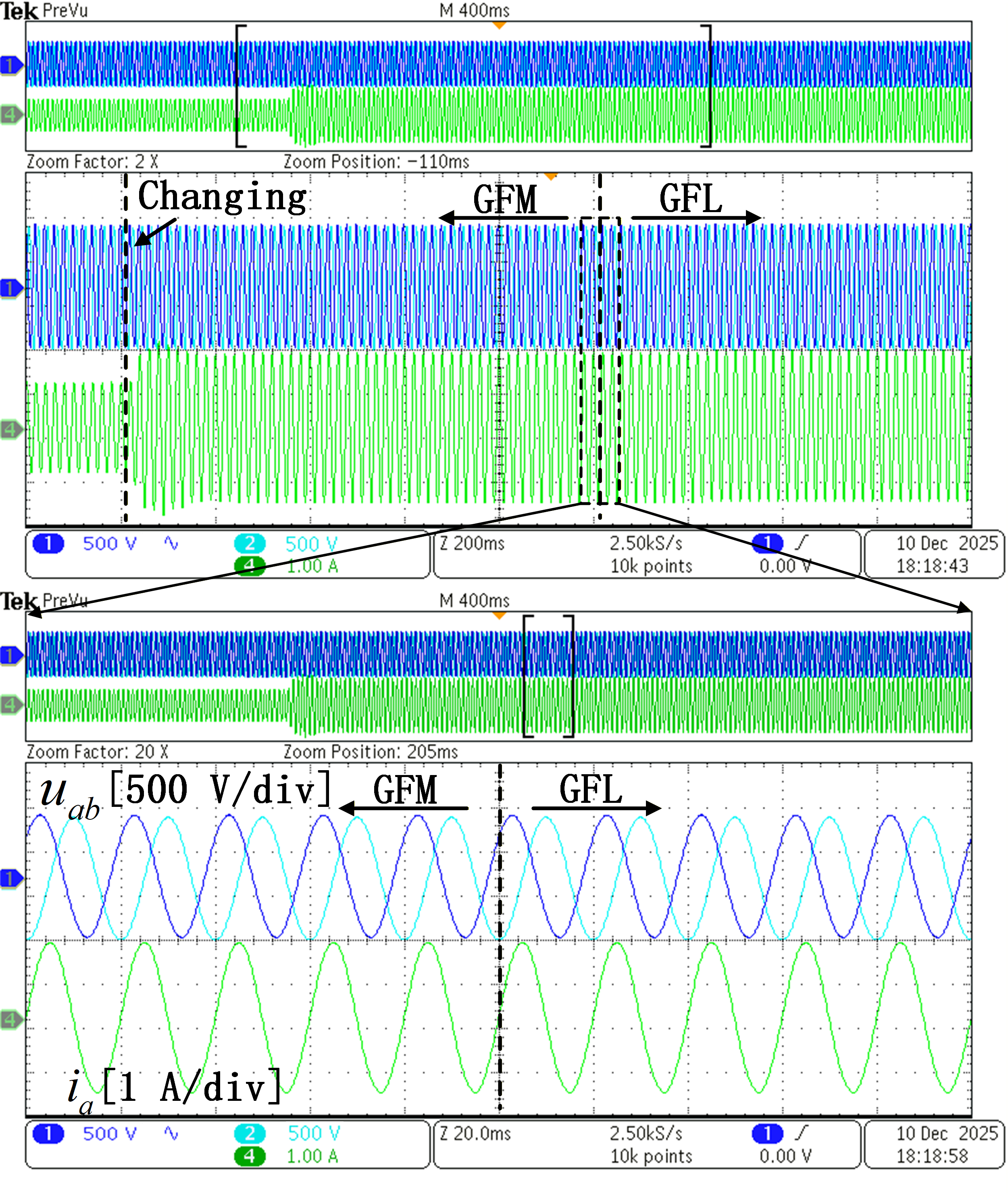}
\caption{Smooth mode transition from grid-forming to grid-following control}
\label{Fig:GFMtoGFL_after}
\end{figure}

\begin{figure}[th!]
\centering
\includegraphics[scale=0.37]{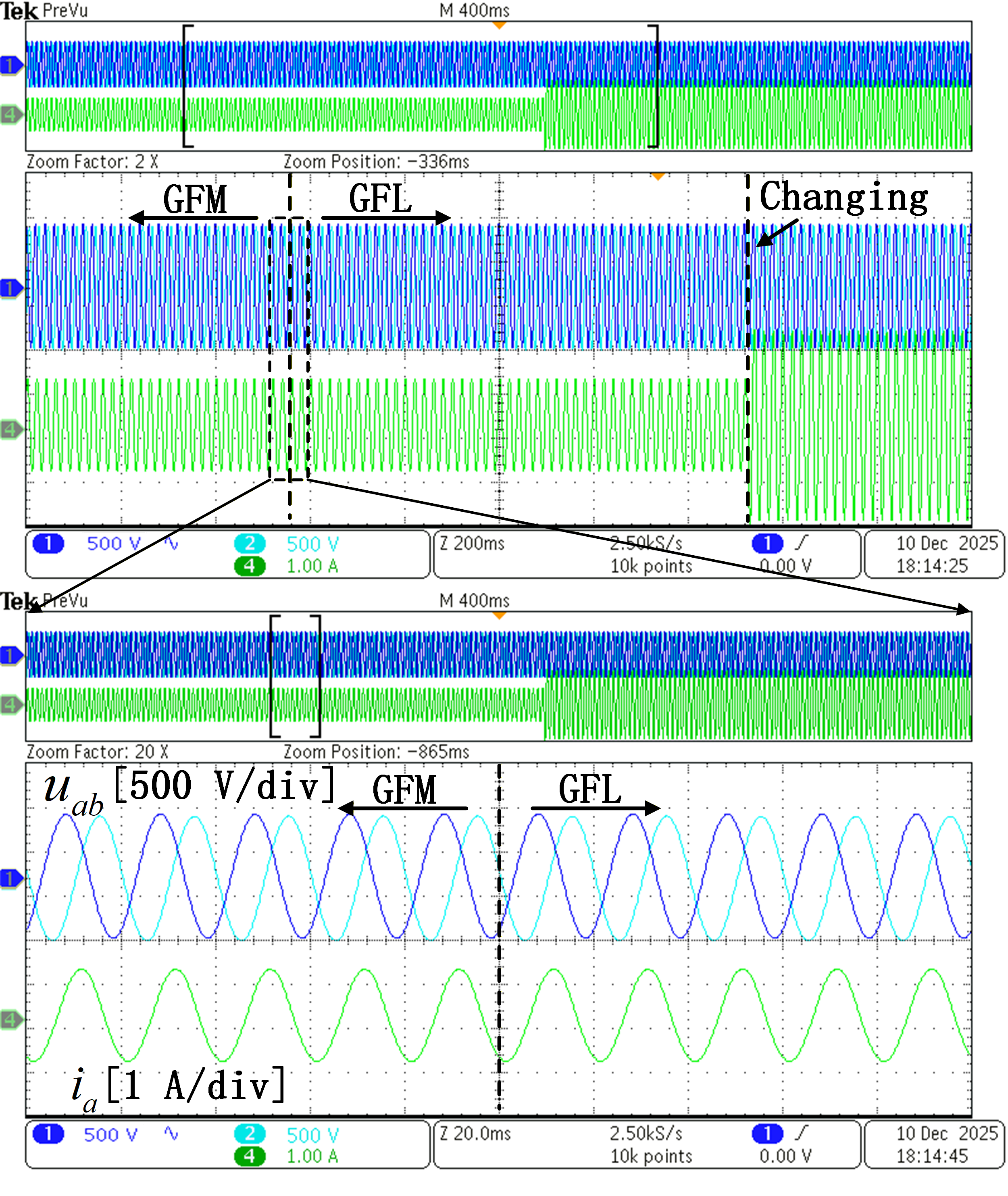}
\caption{Smooth mode transition from grid-forming to grid-following control}
\label{Fig:GFMtoGFL_before}
\end{figure}

\figref{Fig:GFLtoGFM_after} and \figref{Fig:GFLtoGFM_before} demonstrate the process of achieving a smooth transition from grid-following to grid-forming control between different static operating points via two distinct approaches, as referenced earlier. Similarly, \figref{Fig:GFMtoGFL_after} and \figref{Fig:GFMtoGFL_before} depict the two distinct approaches for achieving a smooth transition from grid-forming to grid-following control between different static operating points.

\section{Conclusions} \label{Section:Conclusion}

This paper proposes a state mapping method that is used for analyzing the mode transition between GFM and GFL, and for helping with the switching control design. The state-space expressions of grid-following and grid-forming controls are represented. Meanwhile, the Lyapunov's theorems are used to obtain the equilibrium points and the domain of attraction of the system, thereby modeling the transition process between GFM and GFL control. Controllers for achieving ideally-smooth bidirectional switching between GFM and GFL are designed and verified, through theoretical analysis, simulation modeling, and experiment tests.





\ifCLASSOPTIONcaptionsoff
  \newpage
\fi

\bibliographystyle{IEEEtran}
\bibliography{Paper}

\end{document}